\documentclass[eqsecnum,amsmath,onecolumn,notitlepage,nofootinbib]{revtex4-1}
\usepackage{custom}

\begin{document}

\title{Relativistic Hydrodynamic Fluctuations}
\author{Xin An}
\email{xan2@uic.edu}

\author{G\"{o}k\c{c}e Ba\c{s}ar}
\email{gbasar@uic.edu}

\author{Mikhail Stephanov}
\email{misha@uic.edu}

\author{Ho-Ung Yee}
\email{hyee@uic.edu}
\affiliation{Department of Physics, University of Illinois, Chicago, IL 60607, USA}

\date{\today}

\newpage
\begin{abstract}

  We present a general systematic formalism for describing dynamics of
  fluctuations in an arbitrary relativistic hydrodynamic flow,
  including their feedback (known as long-time hydrodynamic
  tails). The fluctuations are described by two-point equal-time
  correlation functions. We introduce a definition of {\em equal
    time\/} in a situation where the local rest frame is determined by
  the local flow velocity, and a method of taking derivatives and
  Wigner transforms of such equal-time correlation functions, which we
  call {\em confluent}. We find that the equations for
  confluent Wigner functions not only resemble kinetic equations, but that the
  kinetic equation for phonons propagating on an arbitrary background
  nontrivially  matches  the equations for Wigner functions,
  including relativistic inertial and Coriolis forces due to
  acceleration and vorticity of the flow. We also describe the
  procedure of renormalization of short-distance singularities which
  eliminates cutoff dependence, allowing efficient numerical
  implementation of these equations.

\end{abstract}

\maketitle

\section{Introduction}
\label{sec:intro}

%\section{Introduction}

\subsection{Motivation and overview}
\label{sec:motivation-overview}

Hydrodynamics -- the universal theory describing macroscopic motion of
fluids -- hardly needs an introduction. The range of its applications
is extraordinarily wide -- from molecular biology to astrophysics. The
well-established conceptual text-book framework of hydrodynamics
\cite{Landau:2013fluid} has received considerable renewed attention and further
development from various points of view recently. One of the
drivers of the recent interest is the necessity to develop tools for
quantitative analysis of heavy-ion collisions~\cite{Jeon:2015dfa,Romatschke:2017ejr}.  A major ingredient
which is needed is relativistic hydrodynamics with fluctuations. In
many common contexts fluctuations in hydrodynamics could be considered
negligible, such as in truly macroscopic systems with
$\mathcal O(10^{24})$ particle degrees of freedom. Heavy-ion
collisions, however, occupy a ``sweet spot'' in terms of the size:
with $\mathcal O(10^{2-4})$ particle degrees of freedom the relevant
systems are large enough to be treated hydrodynamically but small
enough for fluctuations to be important and directly observable via
event-by-event measurements. In particular, fluctuations are expected
to be further enhanced if the matter created in the collisions is in a
state close to a critical point. In this case fluctuations can serve
as signatures of the critical point
\cite{Stephanov:1998dy,Stephanov:1999zu,Stephanov:2008qz,Stephanov:2011pb}
in the beam energy scan experiments~\cite{Aggarwal:2010cw}.

In the classic Landau-Lifshitz~\cite{Landau:2013stat2} approach to
hydrodynamic fluctuations the local noise due to microscopic degrees
of freedom is introduced into constitutive equations. Generalizing
this formalism to relativistic hydrodynamics and applying it to
relativistically expanding solutions is one of the approaches pursued
in recent
literature~\cite{Kapusta:2011gt,Kapusta:2012zb,Young:2014pka}. The
main drawback of this approach is that practical implementation (e.g.,
for realistic heavy-ion collision simulations) requires introducing
local noise whose amplitude needs to be taken to infinity as the
coarse-graining distance scale (hydrodynamic cell size) is sent to
zero. Nonlinearities lead to divergent noise-induced corrections to
equation of state as well as transport coefficients and make numerical
simulations difficult if not outright infeasible.

An alternative way of describing dynamical effects of fluctuations was
introduced by Andreev in the 1970s who considered evolution of two-point
equal-time correlation functions \cite{Andreev:1978}. This approach has the
advantage of being formulated in terms of {\em deterministic}
equations, avoiding the ``infinite noise'' problem in the
implementation of the stochastic approach. More precisely, the effects
of the ``infinite noise'' can be isolated and absorbed into
``renormalization'' of the equation of state and transport coefficients
in a close analogy with the renormalization in quantum field theories.
Of course, the stochastic and the correlation function (deterministic)
approaches are equivalent and complementary, in ways very similar to
Langevin and Fokker-Plank description of stochastic processes, and the
ultimate choice is to be made based on practicality, in particular,
for numerical simulations.

The deterministic approach (also referred to as hydro-kinetic approach
due to the similarity of some of the additional equations to kinetic equations
for phonons
\footnote{Despite this similarity to kinetic theory, the correlation function approach does not rely on validity of any underlying microscopic kinetic description. As hydrodynamics itself, the approach is applicable for either weakly or strongly coupled quantum field theories. The quasiparticles described by "hydro-kinetic" equations are {\em macroscopic\/} hydrodynamic excitations, such as phonons.
}) has been recently discussed in a
relativistic context~\cite{Akamatsu:2017,Akamatsu:2018,Martinez:2018}
for a special case of Bjorken boost invariant solution
where symmetries allow to reduce the effective dimensionality of the
problem and simplify the analysis.

A more general approach is needed in order to lift the limitations of
the static or boost invariant solution and to enable practical simulations of
relativistic hydrodynamics with fluctuations in a general inhomogeneous
three-dimensional background characteristic of heavy-ion
collisions. Such an approach should, for example, capture the effects
of vorticity, absent in the Bjorken solution, but important in
heavy-ion collisions~\cite{STAR:2017ckg}. The aim of this paper is to
develop such a universal approach.

%%%

\subsection{Variables and scales}
\label{sec:variables-scales}

Hydrodynamic variables are macroscopically averaged values of
densities of  conserved quantities, such as energy and momentum.
The macroscopic averaging is done at fixed time $t$ over a region of
linear size $\ab$ (hydrodynamic cell) around a point with spatial
coordinates $\bm x$. In order to be macroscopic, the length~$\ab$ must
greatly exceed microscopic scales, $\lmic$, such as mean-free path in a
weakly-coupled system, or thermal length~$1/T$ in a relativistic
strongly coupled system 
\begin{equation}
  \label{eq:a-vs-lmic}
 \ab\gg \lmic\,.    
\end{equation}
The resulting coarse-grained
variables can be used to describe evolution of
inhomogeneities at larger length scales
\begin{equation}
L\gg \ab\,.\label{eq:Lb}
\end{equation}

To facilitate the discussion let us refer to the hydrodynamic
variables defined via coarse-graining discussed above as
$\spsi_A(t,\bm x)$, where index $A$ labels a variable. Since we are
describing a thermal system, the variables $\spsi_A$ are stochastic --
fluctuating between members of the statistical ensemble describing our
system (in heavy-ion collisions -- between collision events). To be
more precise, variables $\spsi_A$ are operators. However, due to
macroscopic averaging involved in their construction they behave as
classical (commuting) stochastic variables. Their quantum fluctuations
are negligible compared to (classical) thermal fluctuations.\footnote{The precise condition for that is that the quantum uncertainty of
  the energy due to finite characteristic time of the evolution of
  these variables is much smaller than their typical thermal energy,
  $T$. The fastest evolving degrees of freedom after coarse graining
  are sound modes with wave-length $\ab$. Their frequency $c_s/\ab$
  must therefore be much smaller than $T$, i.e., $\ab\gg c_s/T$. } Due
to coarse graining, fluctuations at scales shorter than~$b$ are
averaged out, i.e., suppressed. In this sense, $\Lambda = 1/\ab$ plays
the role of the ultraviolet (wave-vector) cutoff.

In order to describe fluctuations we introduce the ensemble averages
of the variables $\psi_A \equiv \langle\spsi_A\rangle$. The
ensemble averages~$\psi_A$ obey deterministic hydrodynamic
equations. In addition to these usual hydrodynamic variables
(one-point functions) we must introduce two-point
functions which are ensemble averaged equal-time products at two
space-time points:
$\langle\phi_A(t, \bm
x_1)\phi_B(t, \bm x_2)\rangle$, where
$\phi = \spsi - \langle\psi\rangle$ is the fluctuating part of
the variable, as usual.\footnote{More generally, the variable $\phi_A$
could be (and will be) a linear combination of $\spsi_B-\langle\psi_B\rangle$.}

In equilibrium, the correlators
$\langle\phi_A(t, \bm x_1)\phi_B(t, \bm x_2)\rangle$ are
translationally invariant, i.e., independent of the midpoint
$\bm x \equiv (\bm x_1 + \bm x_2)/2$ at fixed separation
$\bm y\equiv\bm x_1-\bm x_2$. The dependence of correlation functions
of operators in equilibrium on separation $\bm y$ is characterized by
exponential fall-off at distances larger than correlation length
$\xi$: $e^{-|\bm x_1-\bm x_2|/\xi}$.  Correlation length $\xi$ is a
microscopic scale, typically: $\xi\sim \lmic$.~\footnote{Near a
  critical point correlation length is large, e.g., $\xi\gg 1/T$, and
  additional hierarchy of scales emerges. In this case our analysis
  applies if the macroscopic size $\ab$ is taken to be much greater
  than $\xi$: $b\gg\xi$.  Dynamics of fluctuations near a critical
  point in the opposite regime, $\ab\ll\xi$, is characterized by
  dynamical scaling and is discussed in \cite{Hohenberg:1977}.}  From
the point of view of the coarse-grained variables $\phi_A$, therefore,
the distance $\xi\ll \ab$ is negligible and the equilibrium correlator
$\langle\phi_A(t, \bm x_1)\phi_B(t, \bm x_2)\rangle$ is essentially a
multiple of the delta-function $\delta^3(\bm x_1 - \bm x_2)$.

Hydrodynamics, however, describes systems which are not in complete
equilibrium: variations, or gradients, of the variables over
macroscopic scales~$L$ lead to evolution (flow) characterized by time
scale $\tev\sim L/c_s$, where $c_s$ is the sound speed. The
(re)equilibration, as the system evolves, requires transport of
conserved quantities, which is a diffusive process. Therefore
equilibrium can be established only over scales which can be reached
by diffusion over time of
order~$\tev$:
\begin{equation}\label{eq:leqb}
\leqb \sim \sqrt{\gamma\tev}\sim \sqrt{{\gamma L}/{c_s}}\,,
\end{equation}
where $\gamma$ is an appropriate diffusion constant (typically,
$\gamma\sim 1/T$).~\footnote{In terms of the notation $k^*$ from
  Ref.~\cite{Akamatsu:2017}: $\leqb = 1/k^*$.}  This means for
distances $\ab\ll|\bm y|\ll \leqb$ or, more precisely, for
wave-vectors $\bm \p$ conjugate to $\bm y$ such that
$ 1/\leqb\ll|\bm \p|\ll\Lambda$ the equilibration is
complete. However, at scales around
$\p \sim 1/\leqb\sim \sqrt{c_s k/\gamma}$, where $k\sim 1/L$, the
equilibration is ongoing, as it is trying to catch up with the
evolution of the system. It is this competition between the
equilibration and evolution that we will be describing.

\begin{figure}
 \centering
  \includegraphics[height=10em]{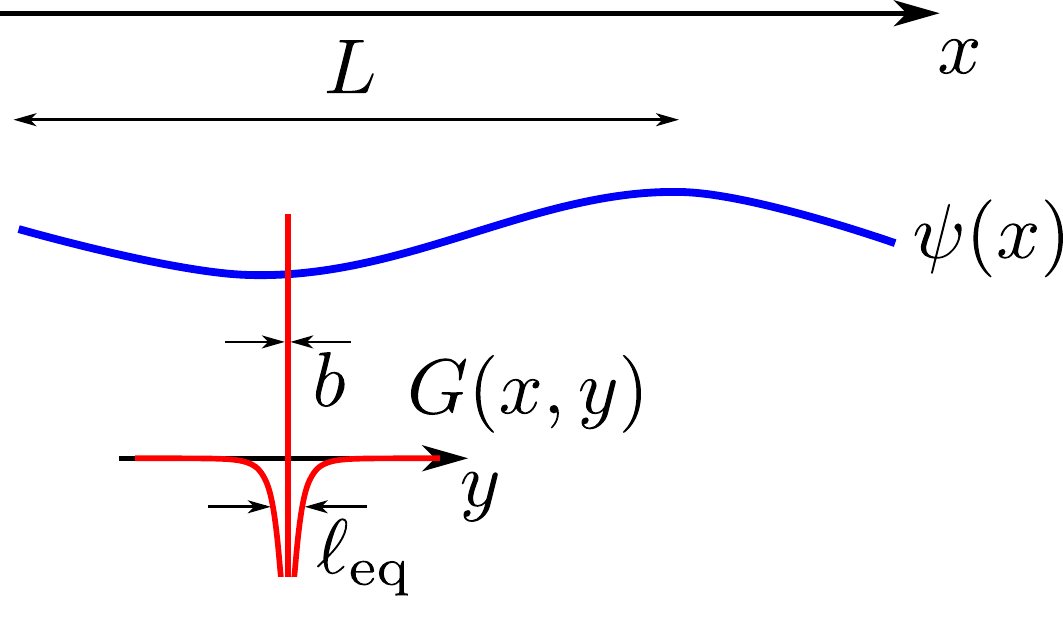}
  \caption{Schematic illustration of various scales described in the
    text. The scale $L$ of the variation of the background
    $\psi(x)$ is the longest in the problem. In equilibrium, the
    fluctuation correlator
    $G(x,y)=\langle\phi(x+y/2)\phi(x-y/2)\rangle$ becomes 
a function (illustrated by a sharp peak on the figure) whose width in $y$ is
    narrower than the shortest hydrodynamic scale -- the coarse graining scale
    $\ab$. If the system evolves, the correlations at scales
    $\leqb\sim \sqrt{\tau_{\rm ev}}\sim \sqrt L$ are not yet
    completely vanishing, giving $G(x,y)$ a finite width of order
    $\leqb\ll L$. A negative contribution from additional
    correlations is necessary to satisfy conservation laws
    $\int_x\phi(x)=0$, i.e., $\int_y G(x,y)=0$ (this integration does
    not commute with $\tev\to\infty$, i.e., equilibrium limit, in
    which the integral is equal to a susceptibility).}
  \label{fig:scales}
\end{figure}

Since the relevant values of $y\sim \leqb$ are
parametrically 
shorter than $L$,
\begin{equation}
  \label{eq:ellL}
  \leqb \sim \sqrt{L/T}\ll L\,,
\end{equation}
we can consider the equilibration process as local,
occurring on a slowly varying background set by local values of
$\psi_A(t,\bm x)$. For that reason it is also naturally convenient to use the mixed
Fourier (i.e., Wigner) transform of the correlation function
$\langle\phi_A(t, \bm x+\bm y/2)\phi_B(t, \bm x-\bm
y/2)\rangle\equiv G_{AB}(x,\bm y)$ with respect to separation
vector $\bm y$,
which we shall denote  $\W_{AB}(x,\bm
\p)$. The relevant values of $\p$ will satisfy
\begin{equation}\label{eq:scales}
k,\gamma \p^2/c_s\ll \p
\ll \Lambda\ll T \,,
\end{equation}
where, for simplicity, we took $\lmic^{-1}\sim T$.
The condition $k\ll \p$ allows us to treat background as smooth when
describing the relaxation of correlations $\W_{AB}( x,\bm \p)$ to
equilibrium. However, we must retain non-zero gradients, $\partial_\mu \psi_A$,
(proportional to $k$) of the background variables
 in the equations for
Wigner functions $\W_{AB}$, since those gradients drive the deviations of
$\W_{AB}$ from equilibrium.

 The fluctuations described by $\W_{AB}$, in turn, feed back
  into constitutive equations which determine the evolution of the
  background flow. One must, therefore, solve the equations for the
  background flow together with the equations for $\W_{AB}$ that we
  are going to derive in this work. 

  Fluctuations with all wave vectors $q$ up to the cutoff $\Lambda$
  contribute to this feedback. The integral of the contributions over
  $q$ is divergent, i.e., it depends polynomially on the cutoff
  $\Lambda$. This would cause difficulties in numerical implementation of
  the hydrodynamic equations and is the manifestation of the
  ``infinite noise'' problem. In a remarkable similarity to the
  renormalizaton of wave-functions and couplings in quantum field
  theories, the fluctuations in hydrodynamics renormalize variables
  (energy density and flow velocity) and parameters (equation of state
  and transport coefficients). The hydrodynamic renormalization
  absorbs the leading large-$q$ terms in $\W(x,q)$ responsible for
  divergences and thus removes the polynomial dependence on the cutoff
  $\Lambda$, allowing efficient numerical implementation.

  Once these cutoff-dependent contributions are absorbed into
  the ``renormalized" hydrodynamics, the true (observable) feedback of the
  out-of-equilibirum fluctuations comes predominantly from modes
  with wave vectors $\p \sim 1/\leqb$. Most importantly, it is
  finite and cutoff independent. The magnitude of these
non-equilibrium effects can  be estimated as the phase-space
volume $\int d^3\p \sim \leqb^{-3}\sim (c_s k/\gamma)^{3/2}$.  The
power $3/2$ indicates that these effects are non-local. They are known
as ``long time tails" of hydrodynamic response
\cite{andreev1970twoliquid,Andreev:1978,Kovtun:2003,Kovtun:2011np}. Their
contribution is typically more important than that of the second order
$\mathcal O(k^2)$ terms in the hydrodynamic derivative expansion
(unless suppressed by a microscopic parameter, such as e.g., number of
colors in a gauge theory)
\footnote{The second order terms are also used in numerical
    applications of hydrodynamics to ensure causality and stability \cite{muller,is}.}.

The discussion of the correlation function above has glossed over an
important issue: ``equal time'' in the definition of $\N_{AB}$ implies
a certain choice of the frame with respect to which equality of time,
i.e., simultaneity, is to be determined. This problem does not arise
in non-relativistic hydrodynamics, but in the case of heavy-ion
collisions it is essential, since the relative velocities at different
points in the fireball are comparable to the speed of light. If the
fluid moves as a whole, with the same velocity (in the lab frame), the
rest frame of such a fluid, not the lab frame, is the natural
choice. In the cases of interest, such as relativistically expanding
fluid, the local rest frame of the fluid is a function of space and
time. We can describe it, as usual, by the 4-velocity $ u(x)$
(macroscopically averaged as described above). 
Therefore, to define the equal-time correlation function
%$\N_{AB}(x;y_\perp)$
we consider correlator
\begin{equation}
\N_{AB}(x,y)\equiv
\av{\dph_A(x^+)\dph_B(x^-)}\label{eq:GAB}
\end{equation}
and evaluate it at points
\begin{equation}
x^\pm = x\pm y/2\label{eq:xpm}
\end{equation}
where 4-vector $y$ lies in the hyperplane orthogonal to $ u(x)$:
$ u(x)\cdot y=0$.~\footnote{In
  Section~\ref{sec:conservative-fluid} we shall discuss the choice of
  the equial-time hypersurface which is not a plane, so as to take
  into account the variation of $ u$ between points $x$ and
  $x_\pm$ and see what, if any, modifications of the results this
  entails. One can anticipate that these modifications will be
  insignificant because the typical range of the correlation function,
  $\sqrt{\gamma L/c_s}$, is short compared to the scale $L$ over which
  the background $ u$ varies significantly.}  The
  corresponding wave vector~$q$ in $\W_{AB}(x,q)$ also resides in 
  a hyperplane orthogonal to $ u(x)$, which is $x$-dependent. We
  find it useful to introduce a type of space-time derivatives which
  account for this $x$-dependence due to inhomogeneous flow and which
  we call ``confluent'' derivatives.

The paper is organized as follows: 
In Section~\ref{sec:stoch-hydr-evol} we introduce stochastic
hydrodynamics and expand its constitutive equations up to quadratic order in fluctuations
around an arbitrary background.
We use linearized hydrodynamic
equations  for fluctuations to derive equations obeyed by two-point correlators.
In Section~\ref{sec:covar-deriv-conn} we introduce confluent
derivative and Wigner function which allow us to write the equations
obeyed by ``equal-time'' correlators. These equations are presented
and studied in
Section~\ref{sec:kinetic}. 

In Section~\ref{sec:diagonalization} we observe that some components
of $W_{AB}$ oscillate at frequencies of order $c_s q$, which are
faster than the evolution of the background and thus, for most
practical purposes, can be averaged out by introducing additional
temporal coarse-graining scale $\ab_t\gg 1/(c_s\p)$. The equations for remaining,
slower components simplify. 

In Section~\ref{sec:renormalization} we consider in detail the
  fluctuation contributions due
  to nonlinearities, and review a general procedure of renormalization
  of first order hydrodynamics. We study the asymptotic behavior of
$\W_{AB}$ at large $q$ and identify the parts of $\W_{AB}$ that
  lead to renormalization of the equation of state and the transport
  coefficients.

In Section~\ref{sec:phonon} we obtain equations of motion for a phonon
in a non-trivial flow
using variational principle, find the corresponding kinetic Liouville
operator, and show that it exactly matches, in several nontrivial
ways, the kinetic equation derived in
Section~\ref{sec:diagonalization}.

Several Appendices contain useful supplementary information. In
particular,
we assemble a
list of our notation choices  used throughout the paper in
Appendix~\ref{sec:notations}.

%%% MS: this is useful for emacs:
%%% Local Variables: 
%%% TeX-PDF-mode: t
%%% TeX-master: "main.tex"
%%% End: 

\section{Stochastic hydrodynamics and fluctuations}
\label{sec:stoch-hydr-evol}
\subsection{Stochastic hydrodynamics}
\label{sec:stochastic-hydro}

Hydrodynamic equations express conservation and transport of energy
and momentum densities:\footnote{For simplicity we do not consider any
  additional conserved charge in this paper. This generalization will
  be addressed in future work.}
\begin{equation}
\del_\mu \sth T^{\mu\nu}=0\label{eq:dT=0}\,.
\end{equation}
To simplify notations later in the paper we label
fluctuating hydrodynamic quantities with an accent, as in $\sth T^{\mu\nu}$, to
distingish them, where necessary,  from quantities which are not fluctuating.
The four conservation equations~(\ref{eq:dT=0}) are solved for the same number of hydrodynamic
variables. A convenient covariant choice for them is the fluid
velocity $\sth u^\mu$ (normalized as $\sth u\cdot \sth u=-1$) and the
energy density $\sth\eps$ in the rest frame of the fluid, that are defined by the Landau's matching condition
\begin{equation}
  \label{eq:Tu=eu}
  -\sth T^{\mu}_\nu \su^\nu = \seps \su^\mu.
\end{equation}
To form a closed system, we need six additional (constitutive)
equations to express all components of $T^{\mu\nu}$ in
terms of $\eps$ and $u^\mu$. For macroscopically large scale dynamics of hydrodynamic variables, $T^{\mu\nu}$ can be expanded in gradients
of $\eps$ and $u^\mu$. The first-order (Landau-Lifshitz) hydrodynamics
corresponds to truncating this expansion at first order in
gradients: 
\begin{eqnarray}\label{eq:Tmunu-constit-0}
{T}^{\mu\nu}(\eps,u)&=&w(\eps) u^\mu u^\nu +p(\eps) g^{\mu\nu}+\Pi^{\mu\nu} ,
\end{eqnarray}
where $p(\eps)$ is pressure as a function of $\eps$ -- also known as
the equation of state and
$ 
w(\eps)=\eps+p(\eps)\,
$ is the enthalpy. The viscous tensor is linear in gradients of $u$:
\begin{eqnarray}
\Pi^{\mu\nu}&=&-2\eta \left(
\h^{\mu\nu}
-{1\over3} \Delta^{\mu\nu}\divu\right)-\zeta\Delta^{\mu\nu}
                \divu,
\label{eq:Pimunu}
\end{eqnarray}
where  shear and bulk viscosities are denoted as $\eta$ and $\zeta$,
respectively, and 
\begin{equation}
\Delta^{\mu\nu}=g^{\mu\nu}+ u^\mu  u^\nu, \label{eq:Delta-definition}
\end{equation}
is the projection operator to the spatial hypersurface orthogonal to $ u$, in terms of which
we define:
\begin{equation}
  \label{eq:theta-def}
  \h^{\mu\nu} = \frac12\left(\partial_\perp^\mu  u^\nu
  + \partial_\perp^\nu  u^\mu\right)\,,
\quad \h=\theta^\mu_\mu\,,
\end{equation}
where \begin{equation}
  \label{eq:partial_perp}
  \partial_{\perp\mu} = \Delta_{\mu}^\nu\partial_\nu.
\end{equation}

 However, the constitutive equations~(\ref{eq:Tmunu-constit-0}) relating 
$T^{\mu\nu}$ and the hydrodynamic variables are valid only on average,
and there exist random local thermal noise $\sS^{\mu\nu}$ which makes Eq.~(\ref{eq:dT=0}) a stochastic
differential equation with
\begin{eqnarray}\label{eq:Tmunu-constit}
\sth{T}^{\mu\nu}&=&
%w(\seps) \su^\mu \su^\nu +p(\seps) g^{\mu\nu}+\sPi^{\mu\nu} 
T^{\mu\nu}(\seps,\su)
+\sS^{\mu\nu},
\end{eqnarray}
The functions of hydrodynamic variables such as $w$, $p$,
$\Pi^{\mu\nu}$, etc.\ in Eq.~(\ref{eq:Tmunu-constit}) are the same as
in Eq.~(\ref{eq:Tmunu-constit-0}) and Eq.~\eqref{eq:Pimunu} but they are evaluated for
fluctuating variables $\sth u$ and $\sth\eps$.

The hydrodynamic variables in $\eqref{eq:Pimunu}$ fluctuate as
  they are driven by the random noise $\sS^{\mu\nu}$, and we need to
  consider statistical ensemble average over these fluctuations for
  any observables on macroscopic scales. We write our stochastic hydrodynamic
variables $\su^\mu$ and $\seps$ as a sum of their averages,
$u\equiv\av{\su}$, $\eps\equiv\av{\seps}$, and linear
fluctuations around them as:
%\footnote{A note on notation: We use both
%  $ \quad$ and $\av{}$ to denote averaging over fluctuations of the
%  independent variables, $\av{\eps}=\beps$, $\av{u}=\bu$. However the
%  meaning of these notations differ when applied to {\em functions} of
%  $\eps$ and $u$. \ms{Can this be simplified? $\eps$ is a function of
%    $\eps$ too. So we can skip 2 sentences above?} I.e., $ f$ denotes the value of the function at
%  averaged values of the arguments. For example, $\eta\equiv\eta(\beps)$,
%  $\zeta\equiv\zeta(\beps)$,
%  $ T^{\mu\nu}\equiv T^{\mu\nu}[\beps,\bu,\eta,\zeta]$ and
%  $\bD^{\mu\nu}\equiv g^{\mu\nu}+\bu^\mu\bu^\nu$. In contrast,
%  $\av{f}$ denotes the average of the function such as
%  $\av{T^{\mu\nu}}\equiv
%  \av{T^{\mu\nu}[u,\eps,\eta(\eps),\zeta(\eps)]}$,
%  $\av{\Pi^{\mu\nu}}\equiv \av{\Pi^{\mu\nu}
%    [u,\eta(\eps),\zeta(\eps)]}$ etc.}
\begin{equation}\label{eq:du-deps}
\su=\bu+\delta u, \,\seps=\beps+\delta\eps\,.
\end{equation}
By definition, the linear fluctuations vanish upon averaging
\begin{equation}
 \av{\delta u}=\av{\delta \eps}=0 \,.
 \end{equation}
 % We shall use the symbol $\sth{\,}$ to denote a fluctuating variable. We also use the same symbol to denote compactly a function whose argument is fluctuating (e.g. $\sPi^{\mu\nu}=\Pi^{\mu\nu}[\seps,\su^\mu]$) when the argument is not explicitly written.
 
These fluctuations are driven by the noise term $\sS^{\mu\nu}$ with
$\av{\sS^{\mu\nu}(x)}=0$, the strength of which is set by the
fluctuation-dissipation theorem\footnote{Due to the
  presence of gradients, the system is slightly out of equilibrium
  and the fluctuation-dissipation relation given in Eq. \eqref{eq:fdt}
  contains corrections proportional to the gradients. However the
  effects of these corrections are higher order (in $k/q$) in the fluctuation expansion as well as the kinetic equation that we discuss in this paper. Therefore we can safely use relation Eq. \eqref{eq:fdt} with $T$ and $\bw$ being functions of $x$ in the remainder of the paper.},
\begin{eqnarray}
\av{\sS^{\mu\nu}(x)\sS^{\lambda\kappa}(x^\prime)}&=& 2T\left[ \eta\, (\bD^{\mu\kappa}\bD^{\nu\lambda}+\bD^{\mu\lambda}\bD^{\nu\kappa})+\left(\zeta-\frac{2}{3} \eta\right) \bD^{\mu\nu}\bD^{\lambda\kappa} \right] \delta^{(4)}(x-x^\prime)\,.
\label{eq:fdt}
\end{eqnarray}
In principle, it is possible to numerically solve the stochastic
equation $\del_\mu \sT^{\mu\nu}=0$ with some coarse-graining, or
wave vector cutoff $\Lambda$,
which regularizes the infinite amplitude of the noise arising from 
the $\delta^{(4)}(x-x^\prime)$ term. However, as we already mentioned in the
introduction, the results would depend sensitively on the
  cutoff $\Lambda$ due to non-linearity of hydrodynamic equations.

We follow an alternative approach, that is, we
include fluctuation contributions to $\av{\sT^{\mu\nu}}$ by expanding $\sT^{\mu\nu}$ to \textit{second order} in
fluctuations.  The fluctuation contributions to $\av{\sT^{\mu\nu}}$ are
given by two-point correlators of the fluctuations, and to describe
their evolution we derive a
separate set of equations.  After proper renormalization that absorbs cutoff dependence into physical parameters, the equation
of motion $\del_\mu\av{\sT^{\mu\nu}}=0$ along with the equations for the
two-point
functions defines a deterministic coupled time evolution of
the averaged variables and of the correlation functions that
can be solved numerically.

Due to non-linearities in the relation between
the variables $(\eps,u)$ and $T^{\mu\nu}$ in Eq.~(\ref{eq:Tmunu-constit-0}), including non-linearities in the
equation of state, such as
\begin{equation}
p(\seps)=p(\beps)+c_s^2 \delta \eps+ {1\over 2}{dc_s^2\over d \beps}\delta \eps^2+\mathcal{O}(\delta\eps^3)\,,
\end{equation}
where $c_s^2=dp(\beps)/d\beps$ is the square of  sound speed,
$\sT^{\mu\nu}$ in Eq.~(\ref{eq:Tmunu-constit})
contains terms that are nonlinear in fluctuations. Expanding
$\sT^{\mu\nu}$ up to second order in fluctuations and taking the average, we have:  
 \begin{eqnarray}
 \label{avTmunu}
   \av{\sth T^{\mu\nu}(x)}   &=&  T^{\mu\nu}(\eps,u)  
+{1\over 2} {d c_s^2 \over d\eps} \bD^{\mu\nu} \av{\delta\eps\,\delta\eps}  +  (1+c_s^2)\big(  \av{\de\, \du^\mu}\bu^\nu+\av{\de\, \du^\nu}\bu^\mu\big)+ \bw \av{\du^\mu\,\du^\nu}\,
\nn
&=& T^{\mu\nu}(\eps,u)+  {\dot c_s\over\bw} \N_{ee}(x) \bD^{\mu\nu} +  {(1+c_s^2)\over c_s\bw}\big(  \N^{e\mu}(x)\bu^\nu+ \bu^\mu\N^{e\nu}(x)\big)+ {1\over \bw} G^{\mu\nu}(x)\,.
\label{eq:avTmunu}
\end{eqnarray}
Note that we neglected the fluctuations of the viscous part
$\Pi^{\mu\nu}$, which are parametrically smaller than the terms
  kept in the above expansion\footnote{ We rely on $\gamma
    \p\sim \p/T \ll 1$, according to Eq.~(\ref{eq:scales}), where $q$
    is the typical wave vector of the fluctuations.}.  In the last line we introduced the collective notation for the fluctuating modes, $\de$ and $\delta u^\mu$: 
\begin{equation}
\phi_A\equiv(\delta e, \dg_\mu)\equiv(c_s \de, \bw \delta u_\mu)
\label{eq:phi_defn}
\end{equation}
where the scalar $\av{\delta e\delta e}$, vector $\av{\delta e\dg^\mu}$, and tensor $\av{\dg^\mu\dg^\nu}$ components of the two-point correlation function are expressed compactly as
\begin{equation}\label{eq:G_AB-phi-phi}
\N_{AB}(x)\equiv\langle\phi_A(x)\phi_B(x)\rangle\,,
\end{equation}
where $A\in (e,0,1,2,3)$\footnote{The mixed index $A\in (e,0,1,2,3)$
  is raised and lowered by the "metric", $\text{diag}(1,-1,1,1,1)$. 
However the object $ u_A$ is not a vector, rather an array that
conveniently combines scalar and vector modes.}. 
In terms of our definition of the correlator $G_{AB}(x,y)$ in Eq.~(\ref{eq:GAB}),
\begin{equation}
  \label{eq:G(x)}
  G_{AB}(x)= G_{AB}(x,0)\,.
\end{equation}
We can express the fluid velocity in the collective notation as well:
\begin{equation}
 \bu_A\equiv(0,\bu_\mu)\,.
 \end{equation}
It should be noted that not all five variables
$\phi_A$ are independent since, due to normalization $\su\cdot \su =-1$, we have a constraint
$ u^A\phi_A=0$. Correspondingly,
\begin{equation}
  \label{eq:orthogonality}
   u^A(x^+) \N_{AB}(x,y) =  \N_{AB}(x,y) u^B(x^-) =0.
\end{equation}
Finally, we define 
  \begin{equation}\label{eq:csdot}
  \dot c_s={\bw \over c_s}{d c_s \over d \beps} ={d\log c_s \over d \log  s}= {1\over 2}T{dc_s^2\over dT},
  \end{equation}
where $ s=\bw/T$ is the average entropy density.

The functions $\N_{AB}(x)$ in Eq.~(\ref{eq:avTmunu}) are, in general, non-local
functionals of the background fields $\eps$ and~$u$. In the next section, we will derive the evolution equation for them by using the linearized hydrodynamics equation of motion for fluctuations.

\subsection{Linearized stochastic equations for fluctuations}
\label{sec:linearized}

In this section we derive the stochastic equation that governs
the dynamics of the linearized fluctuations of the hydrodynamic modes,
$\delta e$ and $\dg^\mu$. This equation is the building block for the evolution 
equation for the two-point function $G_{AB}(x,y)$ and its Wigner
transform, which we call ``kinetic equation".  The energy-momentum
tensor expanded to linear order in fluctuations is given by
\begin{eqnarray}
\sT^{\mu\nu}&\approx& \bw \bu^\mu \bu^\nu +\bp g^{\mu\nu}+\Pi^{\mu\nu}+{1+c_s^2\over c_s}\delta e \,\bu^\mu\bu^\nu+ \bu^\mu\dg^\nu+\bu^\nu\dg^\mu+c_s \delta e g^{\mu\nu}
\nn
&&-\ge (\partial_\perp^{~\mu}\dg^\nu +\partial_\perp^{~\nu}\dg^\mu )-\left(\gz-\frac{2}{3}\ge\right) \bD^{\mu\nu} \del\cdot \dg + \sS^{\mu\nu}\,,
\label{eq:T_1st_ord}
\end{eqnarray}
where 
\begin{equation}
\ge=\eta / \bw \quad\mbox{and}\quad \gz=\zeta/\bw. \label{eq:gammas}
\end{equation}
In this
expansion the first two terms are zero'th order in gradients and the third term
$\Pi^{\mu\nu}$ is of first order or, equivalently, of order $k$. These three terms constitute
the average background value without fluctuation contributions, i.e.,
$T^{\mu\nu}$ in Eq.~(\ref{eq:Tmunu-constit-0}).
The remaining terms are linear in fluctuations.  We consistently
  neglected several terms (e.g., fluctuations of viscosities) that are
  suppressed by either a factor of $k/q\ll 1$ or
  $\gamma k\sim k/T\ll 1$ compared to the terms being kept, according
  to our hierarchy of scales in Eq.~(\ref{eq:scales}) (recall that~$k$
  is the scale of background gradients, and $q$ is the wave-vector of
  fluctuations).
 
The stochastic equation for the linearized modes follows from the energy momentum conservation \mbox{$\del_\mu \sT^{\mu\nu}=0$}, 
\bea
\del_\mu \sT^{\mu\nu}&=&\del_\mu  T^{\mu\nu}+\del_\mu\left({1+c_s^2\over c_s}\delta e \,\bu^\mu\bu^\nu+ \bu^\mu\dg^\nu+\bu^\nu\dg^\mu+c_s \delta e g^{\mu\nu}\right)
\nn
&&
-\ge \partial_\perp^{~2}\dg^\nu -\left(\gz+\frac{1}{3}\ge\right)\partial_\perp^{~\nu} \del\cdot\dg + \del_\mu \sS^{\mu\nu}=0\,,
\label{eq:T_1st_ord2}
\ea  
where we also neglect several terms based on similar considerations discussed above.
By averaging both sides we obtain $\del_\mu  T^{\mu\nu}=0$ to
leading order in fluctuation expansion, which we insert back into
Eq. \eqref{eq:T_1st_ord2} to arrive at a stochastic differential equation for the linearized fluctuations.  

In terms of the notation $\dph_A$ introduced in Eq. \eqref{eq:phi_defn}, the equation for the linearized fluctuations reads:
\be
\bu \cdot \del \dph_A = - \big( \L+\Q+\K \big)_{AB}
\dph^B-{\xi}_A \,,
\label{eq:1st_ord}
\ee
where $\L$, $\Q$, and $\K$ are $5\times5$ matrix operators. The operators $\L$ and $\Q$ are the
ideal and dissipative terms, respectively, $\K$ contains the corrections due to the
first-order gradients of background flow, and ${\xi}$ denotes the
random noise. Explicitly\footnote{It is useful to keep in mind the
  power counting according to our hierarchy of scales in
  Eq.~(\ref{eq:scales}): $\L\sim \p$, $\Q\sim\gamma \p^2$ and $\K\sim k$.}
\begin{eqnarray}
\L&\equiv &\left( \begin{matrix}
      0 & c_s\partial_{\perp\nu} \\
     c_s\partial_{\perp\mu} & 0 \\
   \end{matrix}\right)
   ,\quad 
\Q\equiv \left( \begin{matrix}
      0 & 0\\
     0 & -\ge \bD_{\mu\nu} \partial_\perp^{~2} -( \gz+\frac{1}{3}\ge)\partial_{\perp\mu}\partial_{\perp\nu}\\
   \end{matrix}\right)   
   \nn
\K&\equiv &
\left( \begin{matrix}
      (1+c_s^2+\dot c_s)\bdivu & 2c_s a_\nu  \\
      {1+c_s^2-\dot c_s\over  c_s}a_\mu        &
     -\bu_\mu a_\nu +\partial_{\perp\nu}\bu_\mu
      +\bD_{\mu\nu}\bdivu\\
   \end{matrix}\right)
    ,\quad
  {\xi}\equiv (0, %c_s(\del_\lambda\bu_\kappa) \sS^{\lambda\kappa},
  \bD_{\mu\kappa}\del_\lambda  \sS^{\lambda\kappa})
   \nn
   \label{eq:1st_ord2}
\end{eqnarray}
where  $a_\mu= \bu\cdot \del \bu_\mu$ is the fluid acceleration. Note that $\dot c_s$ terms arise due to space-time variation of $c_s$ via its dependence on $\beps(x)$.

\subsection{Equations of motion for the fluctuation correlators}
\label{sec:eq-two-point}

Having equipped ourselves with the equations of motion for the linearized
fluctuations,  we now derive the evolution equation
for $G_{AB}(x,y)$ with respect to the midpoint variable $x$ at fixed $y$. This equation will be used in  Section~\ref{sec:kinetic} to eventually obtain the evolution equation for its Wigner transform $\W_{AB}(x,q)$. Using the definition of $\N_{AB}(x,y)$
in Eqs.~(\ref{eq:GAB}) and~(\ref{eq:xpm})
and noting that 
\begin{equation}
\del_\mu \N_{AB}(x,y)=\av{(\del^+_\mu\dph_A(x^+))\dph_B(x^-)}+\av{\dph_A(x^+)\del^-_\mu\dph_B(x^-)},
\end{equation}
we apply Eq. \eqref{eq:1st_ord} and keep only the leading order in derivative
expansion (i.e., retaining terms of order $\gamma \p^2$ or $k$, but not $\p k$,
  consistently with Eq.~(\ref{eq:scales})), to obtain
\begin{eqnarray}
\label{eq:kinetic_eq_y}
\bu \cdot \del \N_{AB}(x,y) &=& -\big(\L^{(y)}+\frac{1}{2}\L+\Q^{(y)}+\K+\Y\big)_{AC} \N^C_{\,\,\,B}(x,y) -  \big(-\L^{(y)}+\frac{1}{2}\L+\Q^{(y)}+\K+\Y\big)_{BC} \N_A^{\,\,\,C}(x,y) \nn
&&+\lim_{\delta t\to0}\frac{1}{\delta t}\int_{\bu\cdot x^+}^{\bu\cdot x^+ +\delta t}  \bu\cdot dx^\prime \int_{\bu\cdot x^-}^{\bu\cdot x^-+\delta t}  \bu\cdot dx^{\prime\prime} \av{\xi_A(x^{\prime+})\xi_B(x^{\prime\prime-})}
\nn
&=& -\big(\L^{(y)}+\frac{1}{2}\L+\Q^{(y)}+\K+\Y\big)_{AC} \N^C_{\,\,\,B}(x,y) -  \big(-\L^{(y)}+\frac{1}{2}\L+\Q^{(y)}+\K+\Y\big)_{BC} \N_A^{\,\,\,C}(x,y) \nn
&&+2 T \bw \Q^{(y)}_{AB} \delta^3(\yp),
\nn
\end{eqnarray}
where we converted the independent space-time variables from $(x^+,
x^-)$ to $(x, y)$ by using Eq.~(\ref{eq:xpm}) and used superscripts
 `$(y)$' on the operators to specify that the derivatives involved are to be taken
with respect to $y$ at fixed~$x$. In particular, the operator
\begin{equation}
\label{eq:L_y}
\L^{(y)}\equiv\left( \begin{matrix}
      0 & c_s(x)\partial_{\perp\nu}^{(y)} \\
     c_s(x)\partial_{\perp\mu}^{(y)} & 0 \\
   \end{matrix}\right)
\end{equation}
comes from the conversion of $x^\pm$ derivatives into $x$, $y$ derivatives, and
\begin{equation}
\label{eq:Y}
\Y\equiv \left( \begin{matrix}
      (1-c_s^2)\bD_{\lambda\kappa} & c_s\bu_\nu\bD_{\lambda\kappa} \\
     c_s\bu_\mu\bD_{\lambda\kappa} & \bD_{\mu\nu}\bD_{\lambda\kappa}-c_s^2\bD_{\mu\lambda}\bD_{\nu\kappa} \\
   \end{matrix}\right)\frac{1}{2}y\cdot\partial\bu^\lambda\partial_\perp^{(y)\kappa}-\frac{1}{2}\frac{\dot c_s}{c_s^2}a\cdot y\L^{(y)}
\end{equation}
collects terms proportional to $y$, which resulted from the
$y$-dependence of $\bu(x^\pm)$ and $c_s(x^\pm)$. The last term in
Eq.~(\ref{eq:kinetic_eq_y}) follows from the usual procedure of
stochastic calculus, by keeping the random noise two-point function in
double integrals over the time interval $\delta t$ and using the
correlation given in Eq. \eqref{eq:fdt}. 

In order to convert this equation into an equation for the Wigner
transform $W_{AB}$ of the correlator $G_{AB}$ we need to define the
Wigner transform more carefully than was necessary until now. 
We also
find it necessary to
introduce a concept of derivative adjusted for the boost by flow,
which one can call ``flow-adjusted derivative'' or ``confluent
derivative''. The concept of frame transformation (boost) involved in its
definition bears some resemblance to the parallel transport in
differential geometry and the derivative itself is similar to
covariant derivative.

%%% MS: this is useful for emacs:
%%% Local Variables: 
%%% TeX-PDF-mode: t
%%% TeX-master: "main-v2.tex"
%%% End: 

\section{Confluent derivative, connection and Wigner function }
\label{sec:covar-deriv-conn}

In this section we discuss several ingredients which we need to
translate equation~(\ref{eq:kinetic_eq_y}) into an equation for the
Wigner function. We begin by discussing how to take derivative of an
{\em equal-time\/} correlator in a situation where the concept of
equal time is different in different space-time points.

In Eq.~(\ref{eq:GAB}) we defined the  equal-time correlator of
hydrodynamic variables as a function of the mid-point $x$ and the
separation vector $y$ as
\begin{equation}
  \label{eq:corr-y}
  \N_{AB}(x,y) \equiv \langle\,\phi_A(x+y/2)\,\phi_B(x-y/2)\, \rangle\,.
\end{equation}
where the domain of $y$ is the 3-dimensional plane orthogonal to $
u(x)$, i.e., $y$ is purely spatial in the local rest frame at $x$.

\begin{figure}
  \centering
  \includegraphics[height=10em]{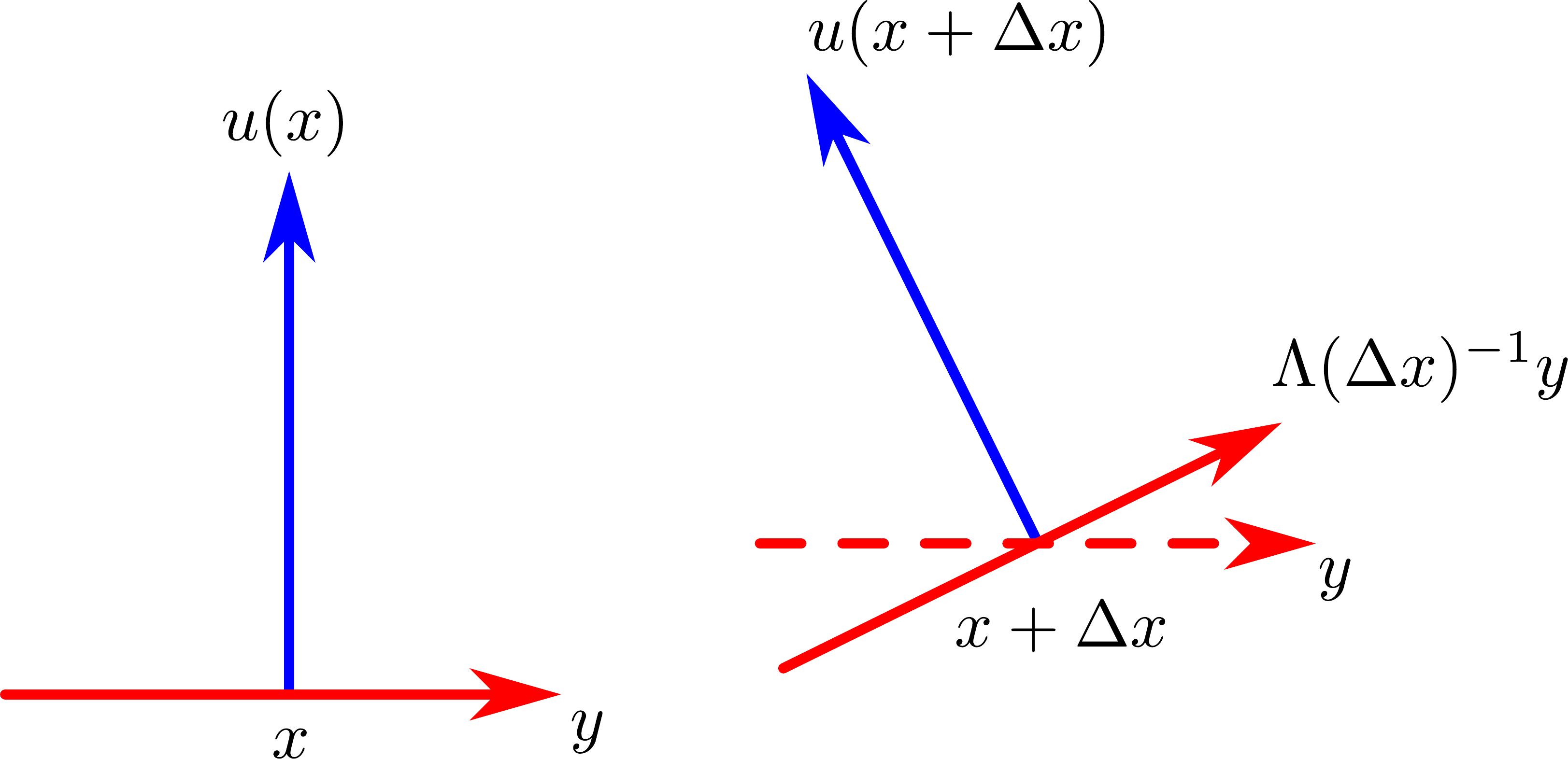}
  \caption{Schematic illustration of the
    Lorentz boost (represented here by an ordinary rotation) of point
    separation vector $y$ needed to keep the point separation purely
    spatial in the local rest frame at a new point $\Delta x$, given
    $ u(x+\Delta x) = \Lambda(\Delta x)^{-1} u(x)$.}
  \label{fig:fixed-y}
\end{figure}

We want to define a partial $x$ derivative of such a function at
``fixed'' $y$. This is not straightforward, as the following expression
illustrates:
\begin{equation}
  \label{eq:dx-nabla-G}
  \Delta x \cdot \partial \N(x,y) = \N(x+\Delta x,y) - \N(x,y)\,.
\end{equation}
In $G(x+\Delta x,y)$ the orthogonality condition
$ u(x+\Delta x)\cdot y =0$ is, in general, false, given
$ u(x)\cdot y=0$ is true: vector $y$ spatial in the frame $ u(x)$
is not spatial in $ u(x+\Delta x)$ (see Fig.~\ref{fig:fixed-y}). To preserve the relationship
between $ u$ and $y$ we need to transform vector $y$ by the same
boost that takes $ u(x)$ to $ u(x+\Delta x)$. Defining this
boost as $\Lambda^{-1}(\Delta x)$ (inverse for later convenience), i.e.:\footnote{Strictly speaking $\Lambda$
  is also a function of $x$ and should be denoted by $\Lambda(\Delta
  x, x)$. For notational simplicity we drop the $x$ argument.}
\begin{equation}
  \label{eq:u-Lambda-u}
  \Lambda(\Delta x) u(x+\Delta x) =  u(x),
\end{equation}
we can then define a derivative at ``fixed'' $y$ as
\begin{equation}
  \label{eq:dx-nabla-G-Lambda-y}
   \Delta x \cdot \cfd G(x,y) = G(x+\Delta x,\Lambda(\Delta x)^{-1}y) -
   G(x,y)\,.
\end{equation}

We have so far suppressed indices $A$ and $B$ in $\N_{AB}$ which label
hydrodynamic variables being correlated. These variables transform
covariantly under Lorentz boosts (five components of $\phi_A$ contain
a scalar and a 4-vector according to Eq.~(\ref{eq:phi_defn})).  It is
natural to define a derivative which measures the
changes of the hydrodynamic variables with respect to the local rest
frame defined by flow velocity $ u$. I.e., we are not interested
in the changes between $\phi_A(x+\Delta x)$ and $\phi_A(x)$ which are
simply due to the difference in the local velocity $ u$, i.e.,
induced by boost transformation from frame $ u(x)$ to
$ u(x+\Delta x)$. In other words, we are interested in
the ``internal'' state of the variables, not affected by frame choice.
The corresponding derivative could be constructed by boosting the
variable $\phi(x+\Delta x)$ in the same way as $ u$ in
Eq.~(\ref{eq:u-Lambda-u}) before comparing to $\phi(x)$,
i.e.,\footnote{Fermi-Walker transport along a world-line is
  constructed in a similar way, in which case $\Delta x$ is
  displacement along the particle's trajectory. In our case $\Delta x$
  can point in any direction, not necessarily along $ u$.}
\begin{equation}
  \label{eq:Dx-nabla-phi}
  \Delta x\cdot\cfd\phi(x) = \Lambda(\Delta x)\phi(x+\Delta x)-\phi(x).
\end{equation} 
With respect to such a derivative, by construction, the flow vector field
$ u(x)$ is  ``constant'':
\begin{equation}
\cfd_\mu u_\nu =0\,,\label{eq:nabla-u=0}
\end{equation}
according to
Eq.~(\ref{eq:u-Lambda-u}). We shall refer to such a derivative as
``confluent'' to distinguish it from a common covariant derivative.

Using explicit form of the infinitesimal boost defined by Eq.~(\ref{eq:u-Lambda-u}):
\begin{equation}
  \label{eq:Lambda-Dx}
  (\Lambda(\Delta x)\phi)_\mu = \phi_\mu 
  -  u_\mu (\Delta  u\cdot\phi) 
  + \Delta u_\mu ( u\cdot\phi)\,,
\end{equation}
where $\Delta u =  u(x+\Delta x) -  u(x)$,
we obtain the explicit expression for the derivative:
\begin{equation}
  \label{eq:nabla-phi}
  \cfd_\lambda\phi_\mu = \partial_\lambda\phi_\mu 
  - \ucon^\nu_{\lambda\mu}\phi_\nu\,,
\end{equation}
where the connection associated with the boost created by flow
gradients is given by
\begin{equation}
  \label{eq:Gamma-udu}
  \ucon^\nu_{\lambda\mu} =  u_\mu\partial_\lambda u^\nu -
   u^\nu\partial_\lambda u_\mu\,.
\end{equation}
Note that this connection is % not symmetric with respect to
% $\lambda\mu$ in contrast to the more familiar metric
% connection, but it is
antisymmetric with respect to $\mu\nu$,
reminiscent of a spin connection. In a sense, it is a spin connection
for a tangent space spanned by hydrodynamic variables $\phi_A$ at point
$x$. In that sense confluent derivative is a covariant derivative for
the connection given by flow gradients in Eq.~(\ref{eq:Gamma-udu}). To
unify equations we can extend the range of indices to
accommodate the full 5-dimensional space of variables and write
\begin{equation}
  \label{eq:nabla-phi-A}
  \cfd_\lambda\phi_A = \partial_\lambda\phi_A 
  - \ucon^B_{\lambda A}\phi_B\,,
\end{equation}
including the case when $A$ or $B$ is $e$. The corresponding
connection is, of course, zero, since $\phi_e=c_s\delta\eps$ is a scalar.

Following the same logic that led us to Eq.~(\ref{eq:Dx-nabla-phi}), we would also like to eliminate the
dependence of the correlator on the difference of the flow velocities
between points $x^+=x+y/2$ and $x^-=x-y/2$. Therefore, we define a confluent
correlation function by boosting both variables $\phi_A(x+y/2)$ and
$\phi_B(x-y/2)$ into the rest frame at the midpoint, $x$, i.e,
\begin{equation}
  \label{eq:LGL}
  \GG_{AB}(x,y) =\,\Lambda^{\phantom AC}_A(y/2)\,\Lambda^{\phantom BD}_B(-y/2)\, \N_{CD}(x,y) \,.
\end{equation}
As a result, the confluent correlator, in contrast to Eq.~(\ref{eq:orthogonality}),
satisfies a simpler orthogonality condition:
\begin{equation}
  \label{eq:GG-u}
   u^A(x) \GG_{AB}(x,y) =  u^B(x) \GG_{AB}(x,y)=0\,.
\end{equation}

Now combining the three ingredients given by
Eqs.~(\ref{eq:dx-nabla-G-Lambda-y}),~(\ref{eq:nabla-phi}) and~(\ref{eq:LGL})  we define the confluent derivative in the following way:
\begin{equation}
  \label{eq:Delta-x-nabla-G-LambdaAB}
     \Delta x \cdot \cfd \GG_{AB}(x,y) = \Lambda(\Delta
     x)^{\phantom{A}C}_A\Lambda(\Delta x)^{\phantom{B}D}_B\GG_{CD}(x+\Delta x,\Lambda(\Delta x)^{-1}y) -
   \GG_{AB}(x,y)\,.
\end{equation}
This expression may be more useful for numerical integration of
equations we derive, where derivatives need to be discretized. The
expression which is used in analytical manipulations is obtained by
Taylor expanding in $\Delta x$:
\begin{equation}
  \label{eq:nabla-mu-G}
  \cfd_\mu \GG_{AB} = \partial_\mu  \GG_{AB} 
- \ucon_{\mu A}^C \GG_{CB} - \ucon_{\mu B}^C \GG_{AC}
-
\econ_{\mu a}^{b}\, y^a \frac{\partial}{\partial y^b} \GG_{AB} \,.
\end{equation}
Another connection, $\econ^b_{\mu a}$ ($a,b=1,2,3$), appears because we need to
define a tangent space at each point $x$ and introduce coordinates,
such as $y^a$, in this space to describe vector $y$ and to keep them
fixed, as we take $x$ derivative (and to take
derivatives with respect to $y^a$ at fixed $x$). To do this we
choose an arbitrary local basis triad, $e_a^\mu(x)$, at each point $x$, such
that $ u\cdot e_a=0$. When we keep vector $y$ ``fixed'' (in the
sense described above), its local coordinates $y^a = e^a_\mu(x) y^\mu$ may
still change due to the rotation of the triad, i.e., not only
$e^a(x+\Delta x) \neq e^a(x)$, but, in general, also $(\Lambda(\Delta x)
e)^a(x+\Delta x) \neq e^a(x)$, in contrast to
Eq.~(\ref{eq:u-Lambda-u}) (see Appendix~\ref{sec:tetrad}).  The last term in Eq.~(\ref{eq:nabla-mu-G})
makes sure this change of the basis is corrected for. In other words, we
need additional connection to  make $e_a^\mu$  confluently constant (like
$ u$ already is without additional connection),~i.e.,
\begin{equation}
  \label{eq:nabla-a}
  \cfd_\lambda e_a^\mu \equiv \partial_\lambda e_a^\mu 
  + \ucon^\mu_{\lambda\nu} e_a^\nu
  -  \econ^c_{\lambda a}e_c^\mu =0\,,
\end{equation}
so that $\cfd_\lambda y = \cfd_\lambda (e_a^\mu(x) y^a)=0$.  The
second term in Eq.~(\ref{eq:nabla-a}) accounts for the boost of $e_a$
as the one in Eq.~(\ref{eq:nabla-phi}) and illustrated in
Fig.~\ref{fig:fixed-y}. In other words, and this is important for
applications, the $x$-deriviative in $\partial_\mu\GG_{AB}$ in
Eq.~(\ref{eq:nabla-mu-G}) is taken at fixed $y^a$ and the boost needed
to keep $y=e_ay^a$ orthogonal to $ u$ is taken care of by
$e_a$. The last term in Eq.~(\ref{eq:nabla-a}) describes the possible
additional rotation of the triad basis vector in the tangent
hyperplane.  This rotation depends on our (arbitrary) choice of the
local triad~$e_a(x)$.

Equation~(\ref{eq:nabla-a}) can be solved for the connection $\econ^a_{\lambda b}$
by multiplying by dual basis vector $e^b_\mu$ such that $e_c\cdot e^b
= \delta_c^b$:
\begin{equation}
  \label{eq:omega-ede}
  \econ^b_{\lambda a} = e^b_\mu \partial_\lambda e^\mu_a \,,
% - e^\mu_b\partial_\lambda e^a_\mu =
\end{equation}
where we used $ u\cdot e^b= u\cdot e_a=0$ and the definition of $\ucon$
connection in Eq.~(\ref{eq:Gamma-udu}). % and $e^a\cdot e_b=\delta^a_b$.
In Appendix~\ref{sec:tetrad}
we provide a simple explicit example of the local triad $e_a$ with the
corresponding connection.

We can now define the Wigner transform of the equal-time correlator
$\GG_{AB}(x,y)$ by integrating over the 3d hyperplane normal to
$ u(x)$ at each point $x$. The integral can be expressed
explicitly as the integral over coordinates $y^a$, in which form it
can be practically evaluated in numerical applications, or, more
formally, as an integral over $y$ constrained to a plane by
$ u\cdot y=0$ condition, i.e.,
\begin{equation}
  \label{eq:int-y}
  \int d^3 y^a = \int d^4y \delta( u\cdot y) \,. %\equiv \int d^3 y
\end{equation}

% Similar to the definition of the covariant derivative, we do not want
% to simply add values of $\N_{AB}(x,y)$ at different points $y$. Instead,
% we boost all $\N_{AB}(x,y)$ into the same point, $y=0$, before adding
% (integrating). 
Thus we arrive at the definition of the confluent
Wigner function:
\begin{equation}
  \label{eq:Wigner-function}
  \W_{AB}(x,\p) \equiv \int d^4y
 \,\delta( u(x)\cdot y)\, e^{-i\p\cdot y}
\, \GG_{AB}(x,y) \,.
%\Lambda^{\phantom AC}_A(y/2)\,\Lambda^{\phantom BD}_B(-y/2)\, \N_{CD}(x,y) \,.
\end{equation}
% We shall omit tilde over $G$ and rely on the second
% argument being $p$ vs $y$ to distinguish Fourier/Wigner transformed
% function. 
Note that due to the delta-function constraint the Wigner function $\W_{AB}(x,\p)$
does not depend on the component of~$\p$ along $ u$
(energy/frequency in local rest frame). Therefore, we only need 3
independent components for vector~$\p$. We shall use the triad basis
we already introduced above for vector $y$ (see also
Appendix~\ref{sec:tetrad}) to express 4-vector $\p_\mu$ in terms
of its 3 independent components $\p_a$ as
\begin{equation}
  \label{eq:qq}
 \p_\mu = e_\mu^a \p_a \,.
\end{equation}

It is now straighforward to write the expresion for the confluent
$x$ derivative of the Wigner function at $\p$ fixed. We need to use the
rules of the Fourier transform to replace $y^a\to i\partial/\partial
\p_a$ and $\partial/\partial y^b\to i\p_b$:
\begin{equation}
  \label{eq:nabla-mu-wigner}
  \cfd_\mu \W_{AB} = \partial_\mu  \W_{AB} 
- \ucon_{\mu A}^C \W_{CB} - \ucon_{\mu B}^C \W_{AC}
+ \econ_{\mu a}^{b}\, \p_b \frac{\partial}{\partial \p_a} \W_{AB} \,,
\end{equation}
where we also took into account antisymmetry of $\econ^a_{\mu b}$. The
partial derivative $\partial_\mu W(x,q)$ is to be taken at fixed $q_a$ (not
fixed $q=e^a(x)q_a$, since that vector has to get boost adjusted to maintain
$q\cdot u=0$).

To simplify notations below we shall also use the following expression involving
derivatives with respect to components of $q$:
\begin{equation}
\label{eq:d/dq}
\frac{\partial}{\partial q_\lambda}
\equiv e^\lambda_a(x)\,\frac{\partial}{\partial q_a}\,.
\end{equation}

%%% MS: this is useful for emacs:
%%% Local Variables: 
%%% TeX-PDF-mode: t
%%% TeX-master: "main.tex"
%%% End: 

\section{Fluctuation kinetic equations}
\label{sec:kinetic}

 The two-point functions $\W_{AB}(x,\p)$ can be viewed as degrees of
freedom additional to the hydrodynamic fields $\psi_A$ (i.e., $\beps$ and $\bu$) in
ways similar to phase-space
distribution functions in kinetic theory.  This
is not just a vague similarity. A certain linear combination of
$\W_{AB}(x,\p)$ can be quantitatively interpreted as phonon
distribution function satisfying Boltzmann equation for a particle
with momentum $\bm \p$ and energy $E=c_s|\bm \p|$ as will be shown in Section~\ref{sec:phonon}. Regardless of this
interpretation, these additional degrees of freedom satisfy a coupled
differential (matrix) equation which we call somewhat loosely the
``fluctuation kinetic equation" or simply "kinetic equation''. The
kinetic equations have to be supplemented by the usual hydrodynamic
field equations of motion (with fluctuation feedback),
$
\del_\mu\av{T^{\mu\nu}}=0 
$,
to obtain a
closed set of equations (somewhat similar to Vlasov equations) to be
solved simultaneously. In this section we derive these
fluctuation kinetic equations, i.e., equations for $W_{AB}$. 

\subsection{Matrix equation for the Wigner function}
\label{sec:kin-eq}

 We return to Eq.~(\ref{eq:kinetic_eq_y}) for $G_{AB}$ and use
it to derive the evolution of the Wigner function defined in the
previous section, expressing all derivatives in terms of the confluent
derivative. Both definitions of the Wigner functions and of the
confluent derivative bring additional terms, but they also lead to
many nontrivial cancellations.  After a rather lengthy and tedious
algebra we find:
\begin{eqnarray}
\bu \cdot \cfd \W(x;\p)&=& -\comm{ i \L^{(\p)}+\K^{(a)}, \W}-  \acomm{{1\over 2} {\Lcon}+\Q^{(\p)}+\K^{(s)}, \W}+\divu \W+2 T w \Q^{(\p)}+(\partial_{\perp\lambda} \bu_\mu) \pp^\mu {\del \W\over \del \pp_{ \lambda}} 
\nn
&& +{1\over2}a_\lambda \acomm{\left(1-\frac{\dot c_s}{c_s^2}\right)\L^{(\p)},  {\del \W\over \del \pp_{ \lambda}}}
+{\del \over \del \pp_{ \lambda}} \left(\{ \Om^{(s)}_\lambda,\W\}+[ \Om^{(a)}_\lambda,\W]-{1\over 4} [\Hmatrix_\lambda,[\L^{(\p)},\W]] \right),
\label{eq:kinetic_eq}
\end{eqnarray}
where 
 \begin{eqnarray}
\L^{(\p)}&\equiv &c_s\left( \begin{matrix}
      0 &  \pp_{\nu} \\
    \pp_{\mu} & 0 \\
   \end{matrix}\right)
   ,\quad 
{\Lcon} \equiv  c_s\left( \begin{matrix}
      0 &  \cfd_{\perp\nu} \\
    \cfd_{\perp\mu} & 0 \\
   \end{matrix}\right)
   ,\quad
\Q^{(\p)}\equiv \left( \begin{matrix}
      0 & 0\\
     0 & \ge \Delta_{\mu\nu} \pp^2 +\left( \gz+\frac{1}{3}\ge\right)\pp_{\mu} \pp_{\nu} \\
   \end{matrix}\right)   
    \nn
      \K^{(s)}&\equiv &
\left( \begin{matrix}
      (1+c_s^2+\dot c_s)\,\divu &  {1\over 2 c_s}(1+2c_s^2)\, a_\nu  \\
 {1\over 2 c_s}(1+2c_s^2)\,a_\mu        &   \Delta_{\mu\nu}\divu+\h_{\mu\nu}\\
   \end{matrix}\right)
,\quad
       \K^{(a)}\equiv 
 \left( \begin{matrix}
     0 &  -   {1-c_s^2-\dot c_s\over  2c_s}a_\nu  \\
  {1-c_s^2-\dot c_s\over  2c_s}a_\mu        & -\omega_{\mu\nu} \\
   \end{matrix}\right)
   \nn
   \Hmatrix_\lambda&\equiv &c_s\left( \begin{matrix}
      0 &  \partial_\nu\bu_\lambda \\
    \partial_\mu\bu_\lambda & 0 \\
   \end{matrix}\right) ,\quad
   \Om^{(s)}_\lambda\equiv {c_s^2\over2}\left( \begin{matrix}
       2\omega_{\kappa\lambda}\pp^\kappa & 0 \\
    0 & \omega_{\mu\lambda} \pp_\nu+ \omega_{\nu\lambda} \pp_\mu  \\
   \end{matrix}\right) ,\quad
    \Om^{(a)}_\lambda\equiv {c_s^2\over2}\left( \begin{matrix}
     0  & 0 \\
    0 & \omega_{\mu\lambda} \pp_\nu- \omega_{\nu\lambda} \pp_\mu  \\
   \end{matrix}\right) 
   \nn
   \label{eq:matrices}
     \end{eqnarray}
with $[A,B]=AB-BA$ and $\{A,B\}=AB+BA$. 
In the expression for
the anti-commutator, the usual matrix multiplication rules are assumed
and the derivates are assumed to act on $\W$. The matrices $\Om^{(s,a)}$ and
$\K^{(s,a)}$ encode the terms proportional to the gradients of the flow
velocity, including the vorticity $\omega_{\mu\nu}$ 
\begin{equation}\label{eq:omega-def}
\omega_{\mu\nu}={1{}\over2} ( \del_{\perp \mu} \bu_\nu-
\del_{\perp\nu} \bu_\mu)\,,
\end{equation}
and its symmetric partner $\h_{\mu\nu}$ defined in Eq.~(\ref{eq:theta-def}).

Note that, within the order of approximation we are working, we can
further use the ideal hydrodynamic equation
$w a_\mu =-\partial_{\perp\mu} \bp $ to eliminate the time-like derivatives of
velocity, i.e., $a_\mu$, on the right-hand side of
Eq.~(\ref{eq:kinetic_eq}). This may be useful for numerical
solution of the equations which would require solving for time
evolution of $ u(x)$ simultaneously.

\subsection{Diagonalization and averaging out fast modes}
\label{sec:diagonalization}

The matrix $\L^{(\p)}$ in the right hand side of the kinetic equation
Eq. \eqref{eq:kinetic_eq} gives the dominant contribution since it is
of order of $\p$ whereas the remaining terms are either order $k$ or
$\gamma \p^2$ both of which are assumed to be much smaller than $\p$
according to our hierarchy of scales in Eq.~(\ref{eq:scales}). Therefore it is useful to express the kinetic equation in the basis where $\L^{(\p)}$ is diagonal. $\L^{(\p)}$ has five eigenvalues: 
\begin{equation}\label{eq:lambda+-}
 \lambda_{\pm}=\pm c_s|\pp|, \quad \lambda_{T_1,T_2}=0, \quad \lambda_\parallel=0\,.
\end{equation}
corresponding to 5 eigenvectors $\psi_{\bf A}$ with ${{\bf A}}=+,-,T_1,T_2,\parallel$. The eigenvectors form a $5\times 5$ matrix
\begin{eqnarray}\label{eq:eigenvectors}
\psi^{\cA}_A=\begin{pmatrix} 
  {1/ \sqrt{2} } &  -{1/ \sqrt{2} } & 0 & 0 & 0 
\\  
 {\php/ \sqrt{2} }  & {  \php/ \sqrt{2} }  & \th^{(1)} & \th^{(2)} & \bu
   \end{pmatrix}\,,
\end{eqnarray}
where $\php =\pp/|\pp|$ is the unit vector along $\pp$ and $\th^{(1)}$ and $\th^{(2)}$ are two transverse unit vectors that satisfy
\begin{equation}\label{eq:ti}
\th^{(i)}\cdot \th^{(j)}=\delta^{ij},\quad \th^{(i)}\cdot\hat \pp=0,\quad \th^{(i)} \cdot \bu=0\,.
\end{equation}
In other words $\th^{(1)}$, $\th^{(2)}$ and $\php$ span the spatial hyperplane orthogonal to $\bu$,
\begin{equation}
\th^{(1)}_\mu \th^{(1)}_\nu+\th^{(2)}_\mu \th^{(2)}_\nu+\php_{\mu} \php_{\nu}=\Delta_{\mu\nu}
\label{eq:tpprojection}
\end{equation} 

The basis vectors in Eq.~(\ref{eq:eigenvectors}) correspond to the
eigenmodes of ideal hydrodynamic equations. Their eigenvalues in Eq.~(\ref{eq:lambda+-})
correspond to positive and negative frequency sound waves and two
degenerate transverse momentum modes. The last zero mode is a
consequence of the orthogonality condition Eq.~(\ref{eq:GG-u}). The
transverse modes are degenerate and the basis in this two-dimensional
subspace can be chosen arbitrarily. 
A convenient explicit choice for $\th^{(i)}$ is given in Appendix
\ref{sec:t_i}. 

We can now transform the kinetic equation~(\ref{eq:kinetic_eq}) into the
diagonal basis of $\L^{(\p)}$ by the orthogonal transformation
$M\rightarrow \psi^T M \psi$~\footnote{Note that since there are
  derivatives with respect to $x$ and $q$ in  Eq. \eqref{eq:kinetic_eq}, one needs to use
  $\psi^T dM \psi= d ( \psi^T M \psi ) 
+[\psi^T d \psi
    , \psi^T M \psi]$.} and express the equation in
terms of new variables:
\begin{equation}
  \label{eq:W_bAbB}
  \W_{\cA\cB}= \psi_{\cA}^A \W_{AB} \psi_{\cB}^B \,.
\end{equation}
The modes
$\W_{\cA \parallel}$, $\W_{\parallel \cB}$, and
$\W_{\parallel \parallel}$ are constrained to vanish by
Eq.~(\ref{eq:GG-u}). We can therefore view $\W_{\cA\cB}$ as
effectively a
$4\times4$ matrix. Furthermore, since in the diagonal basis, 
\begin{equation}
[\L^{(\p)},\W]_{\cA\cB}=(\lambda_{\cA}-\lambda_{\cB})\W_{\cA\cB}, 
\end{equation}
ten of the modes, namely $\W_{\pm \mp}$, $\W_{\pm T_i}$ and
$\W_{T_i\pm}$,
oscillate 
with the frequency of order $c_s \p$, which is much faster than the
background evolution frequency of order $c_s k$. We can use this
separation of time scales to introduce (in addition to spatial coarse
graining at scale $\ab$ described in the Introduction) averaging over
{\em time\/} intervals of order $\ab_t$ such that
\begin{equation}
  \label{eq:at}
  c_s k \ll 1/\ab_t \ll c_s \p\,.
\end{equation}
After such averaging only six components of the matrix $\W$ survive
and equations simplify considerably (as noted in \cite{Akamatsu:2017}).
As a result we are left with six  modes which can be classified into two sound modes $\W_{\pm\pm}$, two transverse modes $\W_{T_1T_1}$ and $\W_{T_2T_2}$, and two shear modes $\W_{T_1T_2}$ and $\W_{T_2T_1}$. The sound modes are completely decoupled and satisfy\footnote{For notational simplicity we denote $\W_{++}$ and $\W_{--}$ simply as $\W_+$ and $\W_-$ respectively.} 
\begin{eqnarray}
\bu\cdot\cfd \W_\pm&=&\mp c_s \php\cdot \cfd \W_\pm - \gL\pp^2(\W_\pm-Tw)
+\left(\pm \left(c_s-\frac{\dot c_s}{c_s}\right)|\pp| a_\mu +(\partial_{\perp\mu} \bu_\nu) \pp^\nu +2 c_s^2 \pp^\lambda \omega_{\lambda\mu}\right){\partial \W_\pm\over\partial \pp_\mu}
\nn
&&-\left((1+c_s^2+\dot c_s)\divu +\h_{\mu\nu}\php^\mu \php^\nu\pm {1+2c_s^2 \over c_s} \php\cdot a \right)\W_\pm\,,
\label{eq:Npm}
\end{eqnarray}
where
\begin{equation}
\gL=\gz+\frac{4}{3}\ge\label{eq:gammaL} \,.
\end{equation}
The confluent derivative of $W_\pm$ is defined as follows: 
\begin{equation}
  \label{eq:nabla-W-def}
  \cfd_\mu \W_\pm \equiv \partial_\mu \W_\pm 
+ \econ^a_{\mu b} q_a\frac{\partial \W_\pm}{\partial q_b}\,.
\end{equation}

The transverse and shear modes mix and satisfy $2\times2$ matrix
equation\footnote{Here $\wtr$ represents the  $2\times 2$ matrix
  $\W_{T_i T_j}$. Similarly, the $ij$ indices of the $2\times2$
  matrices $\btr^{ij}$, $\otr^{ij}$, $\tcon_\mu^{ij}$, $\txcon_\mu^{ij}$ and $\widehat{\mathbb 1}^{ij}=\delta^{ij}$
  are suppressed.} 
\begin{eqnarray}
\bu\cdot \cfd \wtr=-2\pp^2 \ge (\wtr-Tw
  \widehat{\mathbb{1}})
+(\partial_{\perp\mu} \bu_\nu) \pp^\nu
\nabla_{(q)}^\mu\wtr
-\acomm{\btr,\wtr}
+\comm{\otr,\wtr} \,,
\label{eq:Nij}
\end{eqnarray}
where
\begin{eqnarray}
    \btr^{ij}\equiv \frac{1}{2}\divu\,\delta^{ij}+\h^{\mu\nu}\th^{(i)}_\mu \th^{(j)}_\nu,
   \quad\mbox{and}\quad
    \otr^{ij}\equiv \omega^{\mu\nu}\th^{(i)}_\mu\th^{(j)}_\nu, \quad i=1,2;
   \label{eq:ij_matrices}
\end{eqnarray}
and we introduced a covariant $\p$-derivative taking into account rotation
  of the basis $\th^{(i)}(x,q)$ of the transverse modes due to change
  of $q$:
\begin{equation}
    \label{eq:nabla-q}
    \nabla_{(\p)}^\mu\wtr 
    \equiv \frac{\partial\wtr}{\partial{\pp}_\mu}
    + \left[\widehat\omega^\mu,\wtr\right],\quad
    \mbox{where}\quad  \quad \tcon^{ij}_\mu
    \equiv \th_\nu^{(i)}\frac{\partial}{\partial \pp^\mu}\th^{(j)\nu}\,.
\end{equation}

The confluent derivative in Eq.~(\ref{eq:Nij}) also includes
 additional terms associated with $ij$ indices of $\wtr$
 (i.e., of $W_{T_iT_j}$), which are due to the $x$-dependence of
 the basis vectors $t^{(i)}$: 
  \begin{equation}
    \label{eq:nabla-x-tildeW}
        \cfd_\mu\wtr 
    \equiv \partial_\mu \wtr
  + \econ^a_{\mu b}\, q_a \nabla^b_{(q)} \wtr 
     + \left[\txcon_\mu,\wtr\right],\quad
    \mbox{where}\quad  \quad \txcon^{ij}_\mu
    \equiv \th_\nu^{(i)}{\partial_\mu}\th^{(j)\nu}\,.
  \end{equation}
In Appendix~\ref{sec:t_i} we propose a simple and intuitive choice for
the $t^{(i)}$ basis suitable for applications, and compute
corresponding connections $\tcon_\mu^{ij}$ and $\txcon_\mu^{ij}$.

These fluctuation kinetic equations are the central result of
  this work.  By considering these equations together with the
  conservation equation for the energy-momentum tensor, including
  contribution from the fluctuations, such as, e.g.,
  $\N^{\mu\nu}(x)/ w$ in Eq.~(\ref{eq:avTmunu}), we obtain a
  closed system of equations that determines the dynamics of both the
  background flow and the fluctuation correlators
  self-consistently. In order for this program to work in practice, we
  need to deal with the singularity of $\N^{\mu\nu}(x)$ which is
  manifested in the ultraviolet divergence of the the wave-vector
  integral relating $\W_{AB}$ to $\N_{AB}$. To eliminate the resulting
  unphysical cutoff dependence we shall absorb ultra-violet divergent
  contributions of fluctuations into renormalization of a {\it finite}
  number of physical parameters that define first order viscous
  hydrodynamics, i.e. the equation of state and transport coefficients
  (viscosities). The remaining part of fluctuation contributions is
  physical, well-defined and insensitive to the cutoff. In the next
  section, we describe in detail how this renormalization procedure works.

%%% MS: this is useful for emacs:
%%% Local Variables: 
%%% TeX-PDF-mode: t
%%% TeX-master: "main-v2.tex"
%%% End: 

\section{Renormalization of first order hydrodynamics}
\label{sec:renormalization}

\subsection{Short-range singularities and renormalization}
\label{sec:nonlin-renorm}

The ``infinite noise'' that one has to introduce via the
$\delta$-function in Eq.~(\ref{eq:fdt}), which causes
cutoff dependence in solutions of the stochastic
hydrodynamic equations, does have its counterpart in our deterministic
approach. The main advantage of our approach is that it allows us to
{\em analytically\/} separate the effects of the cutoff and to absorb them into
renormalization of hydrodynamic variables $\eps$ and $ u$, as well as the equation of state $p(\eps)$ and
first-order transport coefficients. This procedure has been
discussed by Andreev \cite{Andreev:1978} in non-relativistic
context, and also recently in boost invariant Bjorken background in
Ref.\cite{Akamatsu:2018}. In this section we describe how this
  can be done in relativistic hydrodynamics
in {\em arbitrary\/} background flow.
 
First of all, due to the non-linearity of the energy-momentum tensor in fluctuations, the rest frame defined by the averaged energy-momentum tensor $\langle \sT^{\mu\nu}\rangle$  in Eq.~(\ref{eq:avTmunu})  via Landau's matching
\begin{equation}
  \label{eq:<T>u=eu}
  -\langle \sT^{\mu}_{\nu}\rangle u_R^\nu=\eps_R  u_R^\mu
\end{equation}
is not given by $ u$, i.e., $ u_R\neq  u$ due to fluctuation contributions. Similarly, the energy density $\eps_R$ in the
rest frame $ u_R$ is different from $\eps$.  We shall refer to the hydrodynamic variables $( u_R,\eps_R)$
as ``renormalized'' variables, since they take into accounts the effects
of fluctuations.

The $ u_R$ can be found by first observing that $ u$ can be shifted to eliminate the terms
proportional to $\N^{e\mu}(x)$ in Eq.~(\ref{eq:avTmunu}). 
We also need to keep in mind that due to 
non-linearities in the constraint $\su\cdot \su=-1$, $\bu$ is not properly normalized; namely
\begin{eqnarray}
\bu\cdot\bu=-1-\av{\delta u \cdot \delta u}=-1- \frac{1}{\bw^{2}}\N^\mu_\mu(x)\,.
\end{eqnarray}
Therefore we find the renormalized fluid velocity as\footnote{
Due to our hierarchy of scales in Eq.~\eqref{eq:scales} the fluctuation
  contribution is parametrically small, as we also explain below Eq.~\eqref{eq:p_renorm},
  which justifies the expansion in
  Eq.~\eqref{eq:umu_redefinition}. It
  also means that the difference between $u_R$ and $u$ is negligible
  in subleading (viscous) terms and in kinetic equations such as Eq.~(\ref{eq:kinetic_eq}).}

\begin{eqnarray}
\bu_{R}^\mu \equiv {\bu^\mu+{1+c_s^2 \over c_s\bw^2 } \N^{e\mu}(x)\over \sqrt{1+\N^\mu_\mu(x)/\bw^{2}}}\approx \bu^\mu+{1+c_s^2 \over c_s\bw^2} \N^{e\mu}(x) -{\bu^\mu\over 2 \bw^2 }\N^\nu_\nu(x).
\label{eq:umu_redefinition}
\end{eqnarray}
For notational simplicity we will 
denote $\bu_{R}$ simply as $\bu$ in the following.

Expressing $\av{\sT^{\mu\nu}}$ in Eq.~(\ref{avTmunu}) in terms of
  $\bu_R$ instead of $\bu$ using Eq.~(\ref{eq:umu_redefinition}), substituting into Eq.~(\ref{eq:<T>u=eu}) and multiplying
  both sides by $\bu_{R\mu}$ we obtain  
\begin{eqnarray}
\beps_R\equiv\bw\left(
  1+\frac{1}{\bw^{2}}\N^\mu_\mu(x)\right)-\bp= \beps +{1\over
  \bw} \N^\mu_\mu(x)\,.
\label{eq:eps_redefinition}
\end{eqnarray}
In terms of these renormalized quantities, we have
\begin{eqnarray}
\av{\sT^{\mu\nu}(x)} &=&\beps_R\bu^\mu \bu^\nu + p \bD^{\mu\nu}+\Pi^{\mu\nu}+  {\dot c_s\over\bw} \N_{ee}(x) \bD^{\mu\nu} +{1\over \bw} \N^{\mu\nu}(x)\,.
\label{eq:avTmunu2}
\end{eqnarray}
Note that $ p=p( \eps)$ here is still expressed in terms of ``bare'' energy density.

As usual, due to contribution of short-wavelength fluctuations the coincident point correlators such as $\N_{AB}(x)\equiv\N_{AB}(x,0)$ are divergent, i.e.,
dependent on the wave-vector cutoff $\Lambda$.
These divergences fall into two classes which are easier to disentangle
using the Wigner transform of $\N_{AB}(x,y)$, i.e., Fourier transform
with respect to $y$: $\W_{AB}(x,\p)$ that we define in Section~\ref{sec:covar-deriv-conn}.
In order to study the
short-distance cutoff dependence of $\N_{AB}(x)$ we need to look at
the large-$q$ behavior of $W(x,\p)$,  since
\begin{equation}
  \label{eq:G(x)-W}
  G_{AB}(x) = \int \frac{d^3\p}{(2\pi)^3} W_{AB}(x,\p)\,.
\end{equation}

The leading singularity is apparent even in static homogeneous
equilibrium, since within our coarse-grained resolution
$\N_{AB}^{(0)}(x,y)\sim\delta^3(y)$  (i.e., correlation
length is negligible) and thus $\N_{AB}^{(0)}(x,0)$ is undefined. Of
course, this is an artifact of neglecting the finiteness of
coarse-graining scale $\ab=1/\Lambda$.

The solution of our kinetic equation (\ref{eq:kinetic_eq}) in
  equilibrium is simply given by 
\begin{equation}
\W^{(0)}_{AB}(x, \p)=T\bw \bD_{AB}\,,\label{eq:W0}
\end{equation}
 where
$\bD_{AB}=\text{diag}(1,\bD_{\mu\nu})$. Because $\W^{(0)}$ is
$q$-independent, the integral in Eq.~(\ref{eq:G(x)-W}) is divergent,
i.e., cutoff-dependent:
\begin{eqnarray}
\N^{(0)}_{AB}(x) 
={T w \Lambda^3\over 6\pi^2} \bD_{AB} \,.
\label{eq:Lambda3}
\end{eqnarray}
The corresponding contributions to the energy-momentum tensor in Eq.~(\ref{eq:avTmunu2}) can be absorbed into 
a renormalization of the pressure, i.e., the equation of state.
The renormalized pressure  is then given by
\begin{eqnarray}
p_R\equiv p(\eps) + {\dot c_s\over\bw} \N^{(0)}_{ee}(x)+{1\over 3\bw} {\N^{(0)}}^\mu_\mu(x)\,.\label{eq:p_renorm0}
\end{eqnarray}
Written in terms of the renormalized energy density given in Eq.~(\ref{eq:eps_redefinition}), we obtain the renormalized equation of state as
\begin{eqnarray}\label{eq:p_R}
p_R(\beps_R)= p(\beps_R) + {\dot c_s\over\bw} \N^{(0)}_{ee}(x)+{1-3c_s^2\over 3\bw} {\N^{(0)}}^\mu_\mu(x)= p(\beps_R)+(1-3c_s^2+\dot c_s){T \Lambda^3\over 6\pi^2}\,.
\label{eq:p_renorm}
\end{eqnarray}
It is worth
emphasizing that, even though $\Lambda$ is an ultraviolet cutoff for
the wave-vector of the fluctuating modes, it is still
considered small compared to the microscopic scale, i.e. $\Lambda \ll
T$ (see Eq.~(\ref{eq:scales})). Therefore the ``divergent" contributions of the fluctuations are
still small corrections to the averaged background variables that
are of order $T^4$. However, in numerical simulations, where this
separation of scales in not ideal, these corrections will
introduce noticeable cutoff dependence. Therefore, we would like to
remove these divergent terms not only as a matter of principle, but
also as a practical matter. These considerations are not dissimilar
in  quantum field theories.

In the presence of background gradients, $\W_{AB}(x, \p)$ deviates from the
equilibrium $q$-independent value. This Wigner function is a
solution to an equation we derive in this paper
(Eq.~(\ref{eq:kinetic_eq})) -- a linear differential equation with
coefficients linear in the gradients of velocity. As such, $W_{AB}$ is
a non-local functional of those gradients. The fact that allows us to
remove divergences by redefining physical parameters (as in quantum
field theories) is that the divergent contributions are simply {\em
  local} functions of the velocity gradients. 
 
 Since we are interested in the behavior at large
$q$, responsible for divergences, we shall expand in inverse powers of $q$:
\begin{equation}
\W_{AB}(x, \p)\equiv\W^{(0)}_{AB}(x,\p)+\W^{(1)}_{AB}(x,\p)+ \wt \W_{AB}(x,\p)\,
\label{eq:G_large_p}
\end{equation}
where the first and leading term is the equilibrium value (\ref{eq:W0}).
Using Eq.~(\ref{eq:G(x)-W}) we find correspondingly,
\begin{equation}
  \label{eq:G0G1Gt}
  \N_{AB}(x)\equiv\N^{(0)}_{AB}(x)+\N^{(1)}_{AB}(x)+ \wt \N_{AB}(x)\,.
\end{equation}
The contributions from the first term to the energy-momentum tensor in Eq.~(\ref{eq:avTmunu2}) can be absorbed into 
the renormalization of the equation of state as shown in Eq.~(\ref{eq:p_renorm}).
The second term in the large $\p$ expansion in Eq.~(\ref{eq:G_large_p})
is a {\em local} linear function of
velocity gradients. We shall determine $W^{(1)}_{AB}$ in detail in
the next section. Here we only need to know that it is of order $k/\pp^2$ or, schematically,
\begin{equation}\label{eq:W1}
\W^{(1)}_{AB}(x,\p)\sim {\del \bu\over \gamma \pp^2},
\end{equation}
where $\gamma$ represents relaxation constants proportional to the
viscosities and $\partial u$ represents velocity gradients. Since
only scalar, $\N_{ee}$, and tensor, $\N_{\mu\nu}$, components appear in the expansion
\eqref{eq:avTmunu2}, we focus on those. The phase space
integration in Eq.~(\ref{eq:G(x)-W}) leads to terms linear in
$\Lambda$ which can be decomposed into shear (i.e., traceless) and
bulk viscous terms:
\begin{equation}
{1\over \bw}\N^{(1)}_{\mu\nu}(x)={T \Lambda C_{\rm shear}  \over
  \pi^2} \left(\h_{\mu\nu}
-{1
\over3}\Delta_{\mu\nu}\divu \right) +{T\Lambda C_{\rm bulk}  \over 2 \pi^2} \Delta_{\mu\nu}\divu,\quad {1\over \bw}\N^{(1)}_{ee}(x)={T \Lambda C_{ee} \over 2 \pi^2} \divu\,,
\label{eq:G1}
\end{equation}
where the coefficients
$C_{\rm shear}$, $C_{\rm bulk}$, and $C_{ee}$  are given explicitly in the next section (see Eq.~(\ref{Cvalues})).
Note that only terms satisfying the orthogonality condition
$ u^A(x) G_{AB}(x)=0$ (according to Eq.~\eqref{eq:orthogonality}) can
appear in Eq.~\eqref{eq:G1}.

 The $\mathcal O (\Lambda)$ terms in Eq.~\eqref{eq:avTmunu2} due to 
  $\N^{(1)}_{AB}$ in Eq.~\eqref{eq:G1} can be absorbed by the
  renormalized transport coefficients (namely shear and bulk viscosities) and pressure.  The shear (traceless) term in Eq.~\eqref{eq:G1} can be absorbed by a renormalization of shear viscosity (it has the same form as the shear part of the viscous term $\Pi^{\mu\nu}$ given in Eq.~\eqref{eq:Pimunu}):
 \begin{equation}
 \label{eq:etaren1}
 \eta_{R}=\eta-{T\Lambda C_{\rm shear}  \over 2 \pi^2}.
 \end{equation}
The remaining terms are related to the renormalization of bulk viscosity. To see how this works, let us look at the trace of the energy momentum tensor given in Eq. \eqref{eq:avTmunu2}. Separating the $\Lambda^3$ and $\Lambda$ terms explicitly we have
\begin{eqnarray}
\av{\sT^\mu_\mu(x)}&=&3\left(-\frac{\beps_R}{3}+p(\beps) + {\dot c_s\over\bw} \N^{(0)}_{ee}(x)+{1\over 3\bw} {\N^{(0)}}^\mu_\mu(x)\right)+\left(\Pi^\mu_\mu+{1\over \bw} {\N^{(1)}}^\mu_\mu(x) + 3{\dot c_s\over\bw} \N^{(1)}_{ee}(x)\right)
\nn
&& +3{\dot c_s\over\bw} \wt\N_{ee}(x)+{1\over \bw} \wt\N^\mu_\mu(x),
\label{eq:avTmumu}
\end{eqnarray}
It might be tempting to think that the $\N^{(0)}$ terms that are
independent of flow gradients renormalize the pressure and the
$\N^{(1)}$ terms that are proportional to $\divu$ renormalize the bulk
viscosity. However, this is not entirely correct, because we also have
to take into account that the relation between $\eps$ and $\eps_R$ given by Eq.~\eqref{eq:eps_redefinition} contains
$\N^{(1)}$ terms, which via the renormalized equation of state $p_R(\eps_R)$ in Eq.~\eqref{eq:p_renorm}, contribute to what we mean by
bulk viscous term separated from the renormalized pressure.  Explicitly, 
we first need to express the bare pressure, $p(\beps)$, as a function of the
renormalized $\beps_R$ defined by Eq. \eqref{eq:eps_redefinition},
 \begin{equation}
 p(\beps)=p(\beps_R- \N^\mu_\mu(x)/\bw)\approx p(\beps_R)-{c_s^2\over \bw} \N^\mu_\mu(x)=p(\beps_R)-{c_s^2\over \bw} {\N^{(0)}}^\mu_\mu(x)-{c_s^2\over \bw} {\N^{(1)}}^\mu_\mu(x)-{c_s^2\over \bw} \wt\N^\mu_\mu(x).
 \label{eq:eos_renorm}
 \end{equation}
After inserting this into Eq. \eqref{eq:avTmumu} and using the renormalized equation of state in Eq.~\eqref{eq:p_renorm}, we obtain
\begin{eqnarray}
\av{\sT^\mu_\mu(x)}&=&3\left(-\frac{\beps_R}{3}+p_R(\beps_R)\right)
+\left(\Pi^\mu_\mu+{1-3c_s^2\over \bw} {\N^{(1)}}^\mu_\mu(x) + 3{\dot c_s\over\bw} \N^{(1)}_{ee}(x)\right)\nn
&& +3{\dot c_s\over\bw} \wt\N_{ee}(x)+{1-3c_s^2\over \bw} \wt\N^\mu_\mu(x).
\label{eq:avTmumu2}
\end{eqnarray}
The leading term is the correct ideal part in terms of the renormalized equation of state, and
the $\N^{(1)}$ terms in the
second parentheses in Eq. \eqref{eq:avTmumu2} can now be absorbed by a renormalization of  bulk
viscosity\footnote{In principle, the renormalization of equation of
  state $p(\eps)$ leads to a corresponding renormalization of temperature
  $T(\eps)$. However, since $T$ itself appears only in the noise amplitude
  and therefore only in fluctuation-induced correction, this is a higher order effect.} 
\begin{equation}
\label{eq:zetaren1}
\zeta_R=\zeta-{1-3c_s^2\over 3\bw\divu} {\N^{(1)}}^\mu_\mu(x) - {\dot c_s\over\bw\divu} \N^{(1)}_{ee}(x)=\zeta-{T\Lambda\over 2\pi^2}\left((1-3c_s^2) C_{\rm bulk}+ \dot c_s C_{ee} \right). 
\end{equation}
It is satisfying to see (and is a non-trivial check) that
a conformal symmetry would preserve both the vanishing bulk viscosity $\zeta=0$ and the conformal equation
of state $p=\eps/3$, according to Eq.~\eqref{eq:p_renorm}, under fluctuation corrections, by the virtue of $c_s^2=1/3$.
 
Finally, the last term in Eq. \eqref{eq:G_large_p}, $\wt \W_{AB}(x,\p)$, is
asymptotically of order $k^2/q^4$ and
does not lead to any large-$q$ divergence in $\wt\N_{AB}(x)$ via
integration in Eq.~\eqref{eq:G(x)-W}. Unlike $\N^{(1)}_{AB}(x)$, $\wt\N_{AB}(x)$ is
a {\em non-local\/} functional of velocity gradients and is responsible for the
physical effects known as long-time tails in hydrodynamic response
\cite{Andreev:1978,Kovtun:2003}. These
terms  are finite and constitute the
leading corrections to the hydrodynamic derivative expansion. In terms
of $k$ these corrections
are of noninteger order $k^{3/2}$ which are formally in between the viscous
first-order $\mathcal O(k)$ terms, and the second order
$\mathcal O(k^2)$ terms in constitutive equations.

After the above renormalization of first order viscous hydrodynamics, we finally obtain the
cutoff independent expression for the energy momentum tensor
(constitutive equation):
\begin{eqnarray}
\av{\sT_R^{\mu\nu}(x)} &=&\beps\bu^\mu \bu^\nu +p(\beps) \bD^{\mu\nu}+\Pi^{\mu\nu}+  {1\over \bw} \left(\dot c_s \wt\N_{ee}(x) -c_s^2  \wt\N^\lambda_\lambda(x)\right) \bD^{\mu\nu} +{1\over \bw} \wt\N^{\mu\nu}(x)\,.
\label{eq:avTmunu_ren}
\end{eqnarray}
where we dropped subscript ``R'' on the right hand side. It should be understood that all
  quantities in Eq.~\eqref{eq:avTmunu_ren} and in the kinetic
  equations such as Eq.~\eqref{eq:kinetic_eq} are renormalized. For example,
  the pressure, $p(\beps)$, in Eq.~\eqref{eq:avTmunu_ren} is given by the
  physical equation of state, which could be, e.g., obtained
  from a lattice calculation.
The conservation equations for this tensor
\begin{equation}
  \label{eq:d<TR>=0}
  \partial_\mu\av{\sT_R^{\mu\nu}(x)} =0\,,
\end{equation}
together with the fluctuation kinetic equations~\eqref{eq:Npm}
and~\eqref{eq:Nij} form a closed set of cutoff-independent
hydrodynamic equations.%
\footnote{ The usual concerns about the acausal
  response and associated instabilities in this equation can be
  addressed by the standard Israel-Stewart treatment introducing
  relaxation dynamics for the viscous tensor. This modification
  affects the regime beyond the domain of applicability ($k\ll T$) of
  hydrodynamics \cite{Geroch:1995bx}, and we shall leave it outside the
  scope of this paper, as  part of the set of established procedures
  (such as, e.g., discretization) needed for numerical implementation.}

The necessary components of $\wt\N_{AB}(x)$ can be obtained by solving
the kinetic equations~\eqref{eq:Npm}
and~\eqref{eq:Nij} for $W$  and subtracting the leading large-$q$
contributions $\W^{(0)}$ and $\W^{(1)}$ determined as local
functions of hydrodynamic variables and gradients of velocity by
expressions we shall derive in the next section
(Eqs.~\eqref{eq:Weq},~\eqref{eq:N1pmij},~\eqref{eq:tildeW-W0W1}
and~\eqref{eq:WAB-WcAcB}). Alternatively, with the explicit
expressions for $\W^{(0)}$ and $\W^{(1)}$ given below, one can
substitute \eqref{eq:G_large_p} into Eqs.~\eqref{eq:Npm}
and~\eqref{eq:Nij} and solve the resulting equations
for~$\wt\W$ directly. Using Eq.~\eqref{eq:G(x)-W} we can then determine
\begin{equation}
  \label{eq:Gtilde}
  \wt \N_{AB}(x) = \int \frac{d^3q}{(2\pi)^3}\,\wt W_{AB}(x,q).
\end{equation}

\subsection{Large-$q$ behavior of Wigner
  functions}
\label{sec:lead-subl-large}

As we discussed in the previous section, in order to obtain
finite, cutoff independent equations we need to separate the leading
and subleading large-$q$ terms from the Wigner functions, since these
terms should be absorbed by
renormalization of equation of state and kinetic coefficients.

The leading term is easy to find, since for large $q$ the
$x$-dependence of the background can be neglected and only the
relaxation term $\gamma\pp^2(W-T w)$ in Eqs.~\eqref{eq:Npm}
and~\eqref{eq:Nij} should be kept, with the solution given simply by
local equilibrium values of fluctuations:
  \begin{equation}\label{eq:Weq} 
\W^{(0)}_\pm=T\bw \quad\mbox{ and }\quad \W^{(0)}_{T_i,T_j}=T\bw
  \delta_{ij}.
\end{equation}
This is the consequence of the thermal noise in Eq.~\eqref{eq:fdt}
 satisfying  fluctuation-dissipation theorem. 

In the presence of gradients, the solution deviates from local
  equilibrium at $x$:
\begin{equation}\label{eq:W=W0-DW}
  \W_{\cA\cB}(x,\pp)=\W_{\cA\cB}^{(0)}(x)+\DW_{\cA\cB}(x,\pp)\,.
\end{equation}
  We can substitute this ansatz into Eq.~\eqref{eq:Npm} and~Eq. \eqref{eq:Nij} and use the ideal hydrodynamic equations and
  thermodynamic relations $s=\bw/T$, $dp=sdT$ to evaluate derivatives
  of $\W^{\rm (0)}=T w$ to leading order in flow gradients:
\begin{equation}
\bu\cdot\partial(T\bw)=-(1+2c_s^2)\divu T\bw,\quad \partial_{\perp\mu}(T\bw)=-{1+2c_s^2\over c_s^2} a_\mu T\bw \,.
\end{equation}
As a result we obtain equations for $\DW_{\cA\cB}(x,\pp)$. However, as discussed in the previous section,  $\W_{\cA\cB}$ integrated over $\p$,
still produces an ultraviolet divergence (albeit linear in $\Lambda$ and
not $\Lambda^3$). To isolate this divergence we write
\begin{equation}\label{eq:DeltaW}
\DW_{\cA\cB}(x,\pp)=\W_{\cA\cB}^{(1)}(x,\pp)+\wt\W_{\cA\cB}(x,\pp)\,,
\end{equation}
where we define $\W_{\cA\cB}^{(1)}(x,\pp)$ as the leading term in $\DW_{\cA\cB}(x,\pp)$ in the large $\p$ limit. To find this term we note that
the terms in the equation for  $\DW_{\cA\cB}$ obtained from
Eqs.~\eqref{eq:Npm} and~\eqref{eq:Nij} by substituting Eq.~\eqref{eq:W=W0-DW} fall into
two classes: the terms proportional to $\DW$ (or its derivatives) and the terms independent
of $\DW$. Within each class we identify the leading terms in the
limit of $\p\to\infty$ and require that these terms cancel when we
replace $\DW$ with its leading term, $W^{(1)}$. This gives
us the following equations:
\begin{eqnarray}
0&=&\left((c_s^2-\dot c_s)\divu -\h_{\mu\nu}\hat \pp^\mu \hat \pp^\nu\right)T\bw-\gL\pp^2 \W^{(1)}_\pm\,,
\\
0&=&(1+2c_s^2)\divu T\bw \widehat{{\mathbb 1}}- 2 T\bw \btr-2\ge\pp^2\wtr^{(1)}\,,
\end{eqnarray}
which are easily solved as:
\begin{eqnarray}
\W^{(1)}_\pm(x,\pp)&=& {T\bw\over \gL\pp^2}\left((c_s^2-\dot
                       c_s)\divu -\h_{\mu\nu}
\php^\mu \php^\nu\right)\,,
\nn
\W^{(1)}_{T_iT_j}(x,\pp)&=& {T\bw\over \ge\pp^2}\left(c_s^2\divu\,\delta^{ij}-\h^{\mu\nu}\th^{(i)}_\mu \th^{(j)}_\nu\right)\,.
   \label{eq:N1pmij}
\end{eqnarray}
As expected (see Eq.~(\ref{eq:W1})) these terms are of order
$\partial u/(\gamma \pp^2)$ and lead to order $\Lambda$ terms
after the $\p$ integration in Eq.~(\ref{eq:G(x)-W}). 

The remaining terms in $\DW$ in Eq.~(\ref{eq:DeltaW}), i.e., 
\begin{equation}
  \label{eq:tildeW-W0W1}
  \wt \W = \W-\W^{(0)}-\W^{(1)}\,,
\end{equation}
are of order $k\partial u/(\gamma\pp^2)^2$ at large $\p$ and 
lead to finite $\wt\N(x)$ in Eq.~(\ref{eq:avTmunu_ren}) upon $\p$
integration in Eq.~(\ref{eq:G(x)-W}). 
For finite $\p$ (not satisfying $\gamma\pp^2\gg k$) the dependence of
$\wt\W$ on $\p$ can be represented to linear order in
$\partial u$, schematically, as
\begin{equation}
  \label{eq:tildeW}
  \wt W \sim 
\frac{( u + v)\cdot k}{\gamma\pp^2+i( u + v)\cdot  k}\,
\frac{\partial u}{\gamma\pp^2}
\end{equation}
where $v=\pm c_s\hat\pp$ or $0$ depending on which mode we are
considering. Integration over $d^3\pp$ leads to, also
schematically, $\wt\N(x)\sim k^{1/2}\partial u/\gamma^{3/2}$. The
non-integer power of $k$ represents the fact that $\wt G$ is a
non-local functional of the gradients of $\partial u$. These
non-local terms give rise to power-law (in space and/or time) tails in
hydrodynamic response \cite{Andreev:1978}.\footnote{                          
   Non-linearity in $\partial u$ leads to natural cutoff of the
  power-law tails, i.e., of non-analiticity at
  small~$k$. Schematically, $\wt\N(x)\sim
  (k+\partial u)^{1/2}\partial u$.
}

\subsection{Renormalization of transport coefficients}
\label{sec:renorm-transp-coeff}

Let us now calculate the renormalization of shear and bulk viscosities
using Eq.~(\ref{eq:N1pmij}). First of all, we convert back into the original $e,\mu$ basis in order to plug them into the energy momentum tensor, Eq. \eqref{eq:avTmunu2}. The conversion is given by:
\begin{eqnarray}\label{eq:WAB-WcAcB}
\W_{AB}=\psi^{\cA}_A \W_{\cA\cB} \psi^{\cB}_B 
=\begin{pmatrix} 
{1\over 2}(\W_+ + \W_-) & {1\over 2}(\W_+ - \W_-) \hat \pp_\nu 
\\
{1\over 2}(\W_+ - \W_-)\php_{\mu}  & {1\over 2}(\W_+ + \W_-) \php_{\mu} \php_{\nu} +\W_{T_iT_j}\th_\mu^{(i)}\th_\nu^{(j)}
\end{pmatrix}\,.
\end{eqnarray}
In particular, \eqref{eq:N1pmij} is converted into
\begin{equation}
\begin{gathered}
\W^{(1)}_{ee}(x,\pp)={T\bw\over\gL \pp^2}\left((c_s^2-\dot c_s)\divu -\h_{\mu\nu}\php^\mu \php^\nu \right), \quad \W^{(1)}_{e\mu}(x,\pp)=\W^{(1)}_{\mu e}(x,\pp)=0,\\
\W^{(1)}_{\mu\nu}(x,\pp)={T\bw\over\gL \pp^2}\left((c_s^2-\dot
  c_s)\divu -\h_{\lambda\kappa}
\php^\lambda
  \php^\kappa \right)\php_{\mu} \php_{\nu} +{T\bw\over
  \ge \pp^2}\left(c_s^2 \divu
\,\dtr_{\mu\nu}
  - \h^{\lambda\kappa}\dtr_{\lambda\mu}\dtr_{\kappa\nu} 
  \right),
\end{gathered}
\end{equation}
where $\dtr_{\mu\nu} = \sum_{i=1}^2\th^{(i)}_\mu\th^{(i)}_\nu=\bD_{\mu\nu}-\php_{\mu} \php_{\nu}$.
 With the help of the integrals
\begin{equation}
\begin{gathered}
\int {d^3\pp \over (2\pi)^3} {1\over \pp^2}={\Lambda \over 2\pi^2}, \quad \int {d^3\pp \over (2\pi)^3} {\php_{ \mu} \php_{\nu} \over \pp^2}={\Lambda \over 6\pi^2}\bD_{\mu\nu},
\\
 \int {d^3\pp \over (2\pi)^3} {\php_{\lambda} \php_{\kappa}
   \php_{\mu} \php_{\nu} \over \pp^2}={\Lambda \over 30\pi^2}(\bD_{\lambda\kappa}\bD_{\mu\nu} +\bD_{\lambda\mu}\bD_{\kappa\nu}  +\bD_{\lambda\nu}\bD_{\kappa\mu} ),
\end{gathered}
\end{equation}
we obtain
\begin{equation}
\begin{gathered}
\N^{(1)}_{ee}(x)=-{T\bw\Lambda\over 6\pi^2\gL}\left(1-3c_s^2+3\dot c_s\right)\divu
,\quad \N^{(1)}_{e\mu}(x)=0,\\
\N^{(1)}_{\mu\nu}(x)=-{T\bw\Lambda\over6\pi^2\gL}\left(\left({1\over5}-c_s^2+\dot c_s\right)\divu\bD_{\mu\nu} +{2\over5} \h_{\mu\nu} \right)-{T\bw\Lambda\over 60\pi^2 \ge }\left( 2(1-10c_s^2)\divu\bD_{\mu\nu} +14\h_{\mu\nu} \right),
\end{gathered}
\end{equation}
therefore
\begin{equation}\label{Cvalues}
C_{\rm shear}=-\left(\frac{1}{15\gL}+\frac{7}{30\ge}\right), \quad C_{\rm bulk}=-\left(\frac{1-3c_s^2+3\dot c_s}{9\gL}+\frac{2(1-3c_s^2)}{9\ge}\right), \quad C_{ee}=-\frac{1-3c_s^2+3\dot c_s}{3\gL}.
\end{equation}
Finally inserting these expressions into Eq. \eqref{eq:etaren1} and
  \eqref{eq:zetaren1} we obtain the renormalized shear and bulk
  viscosities
\begin{eqnarray}
\eta_R&=&\eta+{T\Lambda\over 30\pi^2}\left({1\over\gL}+{7\over 2 \ge} \right),\\
\zeta_R&=&\zeta+{T\Lambda\over 18\pi^2}\left({1\over\gL}(1-3c_s^2+3\dot c_s)^2+{2\over\ge}(1-3c_s^2)^2 \right)\,.
\end{eqnarray}
These expressions agree with the results which were computed via
different methods  in the earlier literature (e.g., Eq. (51),
Eq. (A14) and footnote 7 in Ref.~\cite{Akamatsu:2018}). The positivity
of correction to
 $\zeta_R$ is remarkably non-trivial. It follows from
appropriately renormalizing the energy density  as well as
the equation of state  as described by
Eqs. \eqref{eq:p_renorm} and \eqref{eq:eos_renorm}.
It is satisfying to see that the corrections to viscosities are
positive-definite in agreement with the Second Law of Thermodynamics.

%%% MS: this is useful for emacs:
%%% Local Variables: 
%%% TeX-PDF-mode: t
%%% TeX-master: "main.tex"
%%% End: 

\section{Phonon interpretation of the fluctuation kinetic equation}
\label{sec:phonon}

\subsection{Phonon kinetic equation}
\label{sec:phono-eom}

Consider a classical particle whose motion is described in terms of
the space-time vector $x^\mu$ and 4-momentum $\pq^\mu$ with dispersion
relation given by some condition $F(\pq)=0$. For example, for a massive
particle in vacuum $F=\pq^2-m^2$. A phonon dispersion relation is
given by
$p^0=E(\bm\pq)\equiv c_s|\bm\pq|$ in the rest frame of the fluid. This
can be represented by
\begin{equation}
  \label{eq:f(p)}
  F_+(\pq) = \pq\cdot u + E(\pq_\perp),
\end{equation}
where $ u$ is the the 4-velocity of the fluid rest frame $E=c_s|\pq_\perp|$ and
\begin{equation}\label{eq:p_perp}
\pq_\perp^\mu = \pq^\mu + (\pq\cdot  u) u^\mu.
\end{equation}
The classical action can be then written as
\begin{equation}
  \label{eq:action-px}
  S = \int \left(\,\pq\cdot dx - \lambda F_+\,d\tau\,\right)
\end{equation}
where $\lambda$ is a Lagrange multiplier. Variation of the action is given by:
\begin{equation}
  \label{eq:dS-px}
  \delta S  = \int \left[ \delta \pq_\mu\left(dx^\mu 
- \lambda \frac{\partial F_+}{\partial \pq_\mu}d\tau\right)  
+
\delta x^\mu \left(
-d\pq^\mu -  \lambda \frac{\partial F_+}{\partial x^\mu}d\tau
\right)
- \delta\lambda F_+ d\tau
\right]
\end{equation}
Classical trajectory is then given by
equations of motion
\begin{equation}
  \label{eq:var-eom-xdot}
  \dot x^\mu = \frac{\partial F_+}{\partial \pq_\mu}   =  u^\mu + v^\mu 
\end{equation}
where dot denotes $d/(\lambda d\tau)$ (or one can use reparametrization
invariance to set $\lambda\tau$ to equal coordinate time $x^0$ in frame $
u$) and
\begin{equation}
  \label{eq:v}
  v^\mu = \frac{\partial E}{\partial \pq_\mu} 
= \Delta^\mu_\nu\, \frac{\partial E}{\partial \pq_{\perp\nu}}
= c_s\hat\pq_{\perp}^\mu\,,
\end{equation}
(where we used $\partial \pq_{\perp\nu}/\partial \pq_\mu = \Delta^\mu_\nu$) as well as
\begin{equation}
  \label{eq:pdot-eom}
  \dot \pq_\mu = -\frac{\partial F_+}{\partial x^\mu } =
  -\pq_\nu\partial_\mu  u^\nu - \partial_\mu E
\end{equation}
together with the condition $F_+=0$. We
consider local properties of the fluid to be varying
(sufficiently slowly) in space and time. I.e., $ u^\mu=
u^\mu(x)$, as well as $E=E(x, \pq_\perp)$, which in the case of a phonon
means $c_s=c_s(x)$. 

The corresponding Liouville operator acting on 
a function $\mathcal \WN(x,\pq)$ is given by
\begin{equation}
  \label{eq:kin-eq}
\mathcal L[\mathcal\WN]\equiv  \dot x^\mu \frac{\partial \mathcal\WN}{\partial x^\mu} 
+ \dot \pq^\mu \frac{\partial \mathcal\WN}{\partial \pq_\mu}\,.
\end{equation}
Note that $\mathcal L[F_+]=0$. This property is important because it
allows us to restrict the 8-dimensional phase space to the
7-dimensional subspace defined by $F_+=0$, i.e., to consider functions
of the form
\begin{equation}
\mathcal\WN=\delta(F_+) \WN(x,\pq_\perp)  \,,\label{eq:Nhat-N}
\end{equation}
 where $\WN$ is the usual
phase-space distribution function (of 7 variables only). In other
words $\mathcal L[\delta(F_+) \WN] = \delta(F_+)\mathcal L[\WN]$.

In order to write the kinetic equation in terms of the distribution
function $\WN(x,\pq_\perp)$ we need to express $x$ derivatives in
$\mathcal L[N]$ at fixed $\pq$ ($\partial/\partial x^\mu$ in Eq.~(\ref{eq:kin-eq})) in terms of $x$
derivatives at fixed $\pq_\perp$. These derivatives are not the same because the
relationship between $\pq$ and $\pq_\perp$ depends on $x$ (via $ u(x)$
in Eq.~(\ref{eq:p_perp})). One finds
\begin{equation}
  \label{eq:d/dx-nabla}
  \frac{\partial \WN}{\partial x^\mu} = \cfd_\mu \WN 
+ (\partial_\mu \pq_{\perp\nu})\,\frac{\partial \WN}{\partial \pq_{\perp\nu}}\,,
\end{equation}
where we denoted by $\cfd_\mu$ the $x$ derivative at $\pq_\perp$
fixed~\footnote{A more explicit definition involves projections
  $\pq_a$ of $\pq_\perp$ on the local triad
  $\pq_{\perp\mu} = e_\mu^a \pq_a$, in terms of which
  $\cfd_\mu \WN = \partial_\mu \WN + \econ^a_{\mu b} \pq_a\partial
  \WN/\partial \pq_b$ (cf.
  Eqs.~(\ref{eq:nabla-a}),~(\ref{eq:omega-ede})
  and~(\ref{eq:nabla-mu-wigner})). The projections $\pq^a$ are kept
  fixed while taking $x$ derivative, and connection term accounts for
  the rotation of the basis triad $e^a(x)$ which changes $\pq_\perp$
  while $p_a$ is fixed. Similarly, $\pq_\perp$ derivatives at
  fixed~$x$ are more explicitly written as
  $\partial/\partial \pq_{\perp\mu} = e^\mu_a\,\partial/\partial
  \pq_a$ (cf. Eq.~(\ref{eq:d/dq})).}. Corresponingly, the last term in
Eq.~(\ref{eq:pdot-eom}) should be written as
\begin{equation}
  \label{eq:dE}
  \partial_\mu E  = \cfd_\mu E + (\partial_\mu \pq_{\perp\nu})v^\nu.
\end{equation}

Similarly, the $\pq$ derivatives at fixed $x$ should be expressed as
$\pq_\perp$ derivatives
\begin{equation}
  \label{eq:d/dp-d/dp_perp}
  \frac{\partial \WN}{\partial \pq_\mu} = {\Delta}^\mu_\nu \frac{\partial \WN}{\partial \pq_{\perp\nu}}.
\end{equation}
Substituting
Eqs.~(\ref{eq:var-eom-xdot}),~(\ref{eq:pdot-eom}),~(\ref{eq:dE}),~(\ref{eq:d/dx-nabla})
and~(\ref{eq:d/dp-d/dp_perp}) into Eq.~(\ref{eq:kin-eq}) we find
\begin{equation}
  \label{eq:L2}
  \mathcal L[\WN] =
( u + v)\cdot\cfd \WN - 
\left[
\pq_{\perp\nu} \partial_{\perp\mu}  u^\nu + \cfd_{\perp\mu} E + 
v^\nu (\partial_{\perp\mu} \pq_{\perp\nu} - \partial_{\perp\nu} \pq_{\perp\mu})
- ( u\cdot\partial)\pq_{\perp\mu}
\right]\,\frac{\partial \WN}{\partial \pq_{\perp\mu}}\,.
\end{equation}
Finally, using $\partial_\mu \pq_{\perp\nu} = -E \partial_\mu
 u_\nu +  u_\nu \partial_\mu (\pq\cdot u)$, 
we can write the Liouville operator as
\begin{equation}
  \label{eq:L-final}
\mathcal L[\WN] =  ( u+v)\cdot \cfd \WN   
- \left[ E (a_\mu + 2v^\nu\omega_{\nu\mu}) 
+ \pq_{\perp\nu}\partial_{\perp\mu}   u^\nu 
+ \cfd_{\perp\mu} E\right]
\frac{\partial \WN}{\partial \pq_{\perp\mu}}
\end{equation}
The expression in the square brakets is (the negative of) the force
acting on the phonon.~\footnote{One can also obtain this expression by
  taking the spatial projection of the rate of change of $\pq_\perp$,
  i.e., the force is $\Delta_\mu^\nu \dot \pq_{\perp\nu}$, and using
  equations of motion (\ref{eq:var-eom-xdot}) and (\ref{eq:pdot-eom})
  together with Eq.~(\ref{eq:p_perp}).} The two terms in parenthses
multiplied by $E$ are easily recognized as the inertial force due to
acceleration $a$ and the Coriolis force due to rotation
$\omega_{\mu\nu}$, respectively. The force
$-\pq_{\perp\nu}\partial_{\perp\mu}  u^\nu$ is easier to understand by
considering isotropic Hubble-like expansion, i.e., such that
$\partial_{\perp\mu}  u^\nu = H \Delta_\mu^\nu$, where $H$ is the
rate of expansion (Hubble constant). This term then describes the
rescaling of the momentum $\pq_\perp$ (stretching of the sound wave)
due to expansion of the background medium, leading to the ``red
shift'' of the phonon spectrum, similar to the photon red shift in the
expanding universe. 
The last term is the force due to the dependence of energy on the
location of the phonon via the coefficient $c_s$ in its dispersion
relation:
\begin{equation}
  \label{eq:c_s-force}
  -\cfd_{\perp\mu} E = -\frac{\dot c_s}{c_s}|\pq_\perp|a_\mu \,.
\end{equation}

Remarkably, upon changing the notation for the phonon momentum
\begin{equation}
  \label{eq:pq}
  \pq_\perp\to q\,,
\end{equation}
the Liouville operator in Eq.~(\ref{eq:L-final}) with
$E=c_s|\pq_\perp|$ is identical to the one in Eq.~(\ref{eq:Npm})
obtained using completely different (but apparently complementary)
considerations. The two signs in front of $c_s$ in Eq.~(\ref{eq:Npm})
correspond to positive and negative frequency sound waves, or
positive/negative energy solutions of the condition
\begin{equation}
  \label{eq:FF}
  F_+F_-\equiv (\pq\cdot  u)^2 - E^2 =0,
\end{equation}
where $F_\pm = (p\cdot\bu)\pm E$ and the positive energy solution is given by $F_+=0$ in
Eq.~(\ref{eq:f(p)}).

Curiously, for linear dispersion, $E=c_s|p_\perp|$, the condition in
  Eq.~(\ref{eq:FF}) can be written as  $ g^{\mu\nu}\pq_\mu \pq_\nu
  =0$ using flow induced effective ``metric
  tensor'' $ g^{\mu\nu}=-\bu^\mu \bu^\nu +
  c_s^2\Delta^{\mu\nu}$. 
Since
 $d(F_+F_-)=F_-dF_++F_+dF_-$ and $\delta(F)=\delta(F_+)/F_-+\delta(F_-)/F_+$, 
 we see that the equations of motion localized on the $F_+=0$ surface
 are given by
 Eqs.~(\ref{eq:var-eom-xdot}) and (\ref{eq:pdot-eom}) up to rescaling
 of proper time. On the other hand, the equations of motion with the
 constraint $F_+F_-=0$ are given by
 \begin{equation}
 \dot x^\mu=\frac12{\partial (F_+F_-)\over \partial p_\mu}= g^{\mu\nu}p_\nu\quad, \quad\dot p_\mu=-\frac12{\partial (F_+F_-)\over \partial x^\mu}=-{1\over 2}(\partial_\mu
  g^{\alpha\beta})p_\alpha p_\beta\,,
 \end{equation}
from which one can derive the ``geodesic'' equation of motion by taking additional time derivative to the first equation and using these equations once more.
 From this point of view the forces in
  Eq.~(\ref{eq:L-final}) can be viewed as ``gravitational'' forces.

Perhaps even more remarkably than the matching of the Liouville
operators in Eqs.~(\ref{eq:L-final}) and~(\ref{eq:Npm}), the identification
\begin{equation}
\W_\pm(x,q)=c_s|\p|\bw \WN_\pm(x,q)\label{eq:WN}
\end{equation}
leads to nontrivial cancellation of the
whole second line in Eq.~(\ref{eq:Npm}) (i.e., of all terms proportional
to the background gradients $\theta_{\mu\nu}$ and $a^\mu$ times $\W_\pm$) leaving simply the relaxation
term in Eq.~(\ref{eq:Npm}):
\begin{equation}
  \label{eq:N-kinetic}
  \mathcal L_\pm[N_\pm] = -\gL\p^2(N_\pm-T/E)\,,
\end{equation}
where $\mathcal L_\pm$ are different by the sign in front of
$c_s$ in Eq.~(\ref{eq:Npm}).
Note that the equilibrium
value of $N_\pm$,
$\WN_\pm^{(0)}=\W_\pm^{(0)}/(c_s|\pq_\perp|\bw )$, equals $T/E $ as expected for the
low-energy limit of the phonon Bose-Einstein distribution function.

In contrast to Eqs.~(\ref{eq:Npm}) for longitudinal modes which reduces
to a simple form Eq.~(\ref{eq:N-kinetic}) upon rescaling given by
Eq.~(\ref{eq:WN}), Eq.~(\ref{eq:Nij}) for transverse modes cannot
be simplified in this way. This may be related to the fact that there
is no quasiparticle interpretation for these non-propagating,
diffusive modes.

\subsection{Phonon contributions to stress-energy tensor}
\label{sec:phonon-stress}

It is also remarkable that certain contributions of the fluctuations to
stress-energy tensor can be related directly to the
stress-energy tensor of the phonon gas. This provides a
justification to the two-fluid picture (hydrodynamic fluid plus gas of
phonons) which guided the original approach by Andreev~\cite{andreev1970twoliquid}.

Let us start with the expression for the stress tensor for one particle
moving along a trajectory specified by $x(\tau)$:
\begin{equation}
    T_{(1)}^{\mu\nu}(x) = \int d\tau \frac12(p^\mu \dot x^\nu +
p^\nu \dot x^\mu)\delta^4(x-x(\tau)).\label{eq:T1}
\end{equation}
This means for a gas of such particles and holes
 with distribution functions $N_\pm(x,p)$ we have ('s' for 'sound'):
 \begin{equation}
   T_{(s)}^{\mu\nu}(x) = \int_p \frac12(p^\mu \dot x^\nu + p^\nu
\dot x^\mu)(N_+(x,p)-N_-(x,p)).\label{eq:Tmunu_s}
\end{equation}
(Minus here is because a hole is the absence of a particle). Using equation
of motion $\dot x=  u \pm v$ (Eq.~(\ref{eq:var-eom-xdot})) with
$\pm$ for
particles/holes we obtain:
\begin{multline}
  \label{eq:Ts-pEv}
  T_{(s)}^{\mu\nu} = \int_p \Big[E  u^\mu  u^\nu(N_++N_-)
+ \frac12\left((p^\mu + E v^\mu ) u^\nu +
  (\mu\leftrightarrow\nu)\right) (N_+-N_-)
\\+\frac12(p_\perp^\mu v^\nu +(\mu\leftrightarrow\nu) )(N_++N_-)
\Big]
\end{multline}
Using now $E=c_s|p_\perp|$ for the phonon, we find:
\begin{multline}
  \label{eq:Ts-cs}
  T_{(s)}^{\mu\nu} = \int_p \Big[c_s|p_\perp|  u^\mu  u^\nu(N_++N_-)
+ \frac{1+c_s^2}2\left(p^\mu u^\nu +
  (\mu\leftrightarrow\nu)\right) (N_+-N_-)
\\+\frac{c_s}2|p_\perp|(\hat p_\perp^\mu \hat p_\perp^\nu +(\mu\leftrightarrow\nu) )(N_++N_-)
\Big]
\end{multline}
Comparing to Eq.~(\ref{eq:WAB-WcAcB})  (neglecting transverse modes)
and  identifying $N_\pm = \W_\pm/( c_s|p_\perp|w)$
(as in Eq.~(\ref{eq:WN})) we can write:
\begin{equation}
  \label{eq:Ts-G}
  T_{(s)}^{\mu\nu} = \frac1w G^{\mu}_\mu(x) u^\mu  u^\nu 
+ \frac{1+c_s^2}{wc_s}(G^{e\mu}(x)u^\nu+(\mu\leftrightarrow\nu))
+ \frac1w G^{\mu\nu}(x)
\end{equation}
We can recognize the last two terms as the last two terms in
Eq.~(\ref{avTmunu}). The first term contributes to the
shift/renormalization of the energy density
(see Eq.~(\ref{eq:eps_redefinition})).

Phonons contribute to pressure, i.e., renormalize equation of state.
The equilibrium pressure of the phonon gas equals
\begin{equation}
  \label{eq:p_s}
  p_{(s)}=\frac13\Delta_{\mu\nu}T_{(s)}^{\mu\nu}=\frac1{3}\int_pc_s|p_\perp|(N^{(0)}_++N_-^{(0)})
  = \frac1{3w}{\N^{(0)}}^\mu_\mu(x)\,.
\end{equation}
Together with the contribution from the shift of the energy due to
phonons, $-c_s^2{\N^{(0)}}^\mu_\mu(x)/w$, this reproduces the last term
before the second equal sign in Eq.~(\ref{eq:p_R}).

%%% MS: this is useful for emacs:
%%% Local Variables: 
%%% TeX-PDF-mode: t
%%% TeX-master: "main.tex"
%%% End: 

\section{Conclusions and outlook}
\label{sec:conclusions}

We have derived a set of equations which describe coupled evolution of
hydrodynamic variables $\eps$ and $u$ and the correlation
functions of their fluctuations, collectively denoted by $\phi_A$. The
correlation functions are expressed as Wigner transform $W_{AB}(x,\p)$ of
the equal-time correlator $\langle\phi_A\phi_B\rangle$.

An essential feature of this approach, which distinguishes it from the
{\em stochastic\/} approach also pursued in the literature, is the possibility
to cleanly separate the short-range singularities due to ``infinite
noise'' and absorb them into renormalization of the variables,
equation of state and transport coefficients. The success of this
procedure relies on the locality of the short-range singularities in
the same way as in quantum field theoretical renormalization due to
ultraviolet singularities. The resulting constituitive equations for
stress-energy tensor Eq.~(\ref{eq:avTmunu_ren}) contain only {\em finite\/}
contributions from fluctuations.

One of the crucial new issues we tackled is the treatment of the
``equal-time'' in the definition of the correlators. Frame-dependence
of simultaneity is a quintessential relativistic effect and has not
been an issue in the earlier work \cite{Andreev:1978} where similar equations
have been considered in non-relativistic context. Our analysis led us
to a definition of correlators, Wigner functions and $x$-derivatives
adjusted for the changing local rest frame of the fluid. We refer to
the objects which account for the flow in this way as {\em confluent}.

The success of our approach relies significantly on the separation of
scales inherent in the hydrodynamic regime. The scales of homogeneity,
$L$, must be longer (equivalently, the corresponding wave vector
$k=1/L$, must be softer)  than the range of the correlations. In
hydrodynamics, this range, $\leqb$, is of order $\sqrt{\gamma L/c_s}\ll L$. The
corresponding ``hard'' momentum $\p\sim 1/\leqb$ is much larger than the ``soft''
momentum $k$.

Given that this separation of scales is similar to the separation of
scales which leads to kinetic regime in weakly-coupled quantum
field theories, it is
not a coincidence that the equations for the Wigner functions we
obtain are similar to kinetic equations. It is, nevertheless,
remarkable that the equations we obtain by focusing on the
longitudinal mode fluctuations completely coincide with kinetic
equations describing phonon gas on a background with arbitrary non-uniform flow. This includes
nontrivial inertial, Coriolis, and ``Hubble'' forces -- Eq.~(\ref{eq:Npm}) vs Eq.~(\ref{eq:L-final}).
% (which are purely relativistic
% effects, since they are suppressed by the ratio of the phonon speed $c_s$ to
% speed of light, 1, if the phonons are slow).

The originial approach by Andreev postulated Hamiltonian kinetic
equations for a phonon distribution function without
derivation~\cite{Andreev:1978}. We derive these equations
directly from stochastic hydrodynamics and confirm that the
collision/relaxation term has the simplest form assumed in
Ref.~\cite{Andreev:1978} and does not depend on
the gradients of the flow. This result emerges after nontrivial
cancellations (Eq.~(\ref{eq:N-kinetic}) vs Eq.~(\ref{eq:Npm})), which
appear even more nontrivial given that, in
contrast, the ``kinetic'' equation for the transverse modes does
contain flow gradients among the relaxation terms (Eq.~(\ref{eq:Nij})). These
gradient terms were assumed to be absent in Ref.~\cite{Andreev:1978}. We
plan to explore possible implications of these gradient-dependent
relaxation terms in the future.

Although it would be interesting to study possible analytical solutions of our
equations and their consequences, we also hope that these equations
will find application in numerical simulations of the evolution of
heavy-ion collisions or other relativistic many-body systems where
fluctuations are important. Our equations can be directly simulated
for any flow, not necessarily limited by Bjorken boost-invariance
assumption (as in, e.g., Ref.~\cite{Akamatsu:2017}). In particular, the effects of vorticity, absent in the
Bjorken flow, but important in heavy-ion collisions \cite{STAR:2017ckg}, can be studied.

The approach based on the evolution of correlation functions has been
also introduced recently to describe the dynamics near a critical
point \cite{Stephanov:2018hydro+}. The extension of hydrodynamics, or
Hydro+, by a slow mode describing evolution of critical fluctuations
towards equilibrium is a particular example of the correlation
function (hydro-kinetic) approach to hydrodynamic fluctuations. It
would be interesting to re-derive Hydro+ formalism in the framework
used in this paper. In order to do this we must generalize the formalism
to include conserved current (baryon current in QCD), which we defer
to future work~\footnote{Special cases of static and boost
  invariant backgrounds for a conformal
  fluid with conserved
  charge have been discussed in Ref.~\cite{Martinez:2018}.}.  Such
a generalization in essential for the hydrodynamic modeling of fluctuations
and its effects in the the beam-energy scan program~\cite{Aggarwal:2010cw} conducted at RHIC.

Finally, recent advances in formulating hydrodynamics as an effective
field theory on Schwinger-Keldysh path-integration contour, in
principle, allows using powerful field-theoretical methods and
insights, including a diagrammatic machinery for calculating real-time
correlation functions (see, e.g.,
Refs.~\cite{Crossley:2015evo,Haehl:2018lcu} for reviews).  However,
the practical usage is so far mostly limited to correlation functions
in simple backgrounds, such as, e.g., static equilibrium, and it is
not yet clear how to apply this approach to a realistic heavy-ion
collision simulation. It would be interesting to establish an explicit
connection between the formalism we present here and the
Schwinger-Keldysh effective field theory. This could provide better
understanding of some conceptual issues, such as renormalisation at
higher (or even all) orders in hydrodynamic expansion, and help
generalize the approach to tackle higher-order correlation functions
(beyond two-point functions discussed in this paper.).

%%% MS: this is useful for emacs:
%%% Local Variables:
%%% Mode: latex
%%% TeX-PDF-mode: t
%%% TeX-master: "main.tex"
%%% End: 

\begin{acknowledgments}
   We thank Boris Spivak for drawing our attention to Ref.~\cite{andreev1970twoliquid}. 
  We thank Mauricio Martinez and Derek Teaney for helpful discussions.
  This work is supported by the U.S. Department of Energy, Office of
  Science, Office of Nuclear Physics, within the framework of the Beam
  Energy Scan Theory (BEST) Topical Collaboration and grant
  Nos. DE-FG0201ER41195 and DE-SC0018209.
\end{acknowledgments}

\appendix

\section{A local confluent triad}
\label{sec:tetrad}
In order to describe the separation vector $y$ (for example, to enable
numerical solution of the equations for fluctuation correlators) we
need to introduce a basis triad $e^a_\mu(x)$ for the tangent plane
orthoginal to $ u(x)$ at each point $x$. The basis is arbitrary
and here we shall propose a simple and intuitive choice of
$e_a(x)$. We choose a (lab) frame $\mathring u$ and a fixed triad
($a=1,2,3$) satisfying $\mathring e_a\cdot\mathring e^b=\delta^b_a$
and
$\mathring e_a\cdot\mathring u = \mathring e^b\cdot\mathring u
=0$. For simplicity we shall consider an orthogonal triad, equivalent
to its dual, $e^a=e_a$.

 We can then define $e^a(x)$ by a
{\em finite\/} boost from $\mathring u$ to $ u(x)$. The resulting
triad vectors at point $x$ are given by explicit algebraic formulas:
\begin{equation}
e_a = \mathring e_a + ( u + \mathring u)\frac{ u\cdot\mathring e_a}{1- u\cdot\mathring u}\,.
\label{eq:e-o-e}
\end{equation}
One can check that $e_a\cdot u=0$ and $e^a\cdot e_b=\delta^a_b$.

Corresponding spin connection is given by Eq.~(\ref{eq:omega-ede})
\begin{equation}
  \label{eq:omega}
  \econ_{\mu a}^{b} \equiv  e^b_\nu\partial_\mu e_a^\nu = e^b_\nu e_a^\lambda
  \left[\mathring u^\nu\partial_\mu u_\lambda 
    - \mathring u_\lambda\partial_\mu u^\nu\right]
(1- u\cdot\mathring u)^{-1}\,.
\end{equation}

In terms of the confluent connection defined in Eq.~(\ref{eq:Gamma-udu})
one can express spin connection as
\begin{equation}
  \label{eq:omega-Gamma}
  \econ_{\mu a}^{b} 
=  \frac{\ucon^\nu_{\mu\lambda} \mathring e^b_\nu\mathring e^\lambda_a}
{1- u\cdot\mathring  u}\,.
\end{equation}

For certain flow configurations $ u(x)$ it may be possible to find a choice of
triad fields $e_a(x)$ which makes the spin connection $\econ$ vanish. This
requires integrability of Eq.~(\ref{eq:nabla-a}) with $\econ=0$, which
means that the change of vector $e_a$ obtained by integrating
Eq.~(\ref{eq:nabla-a}) with $\econ=0$ between two points should not depend on the path, i.e.,
\begin{equation}
  \label{eq:oint}
  \oint dx^\lambda\ucon_{\lambda\nu}^\mu e_a^\nu=0.
\end{equation}
Using Stokes theorem we see that this is possible if curvature
associated with connection $\ucon^\mu_{\lambda\nu}$ vanishes. Using Eq.~(\ref{eq:Gamma-udu}), we
find:
\begin{equation}
  \label{eq:R}
  \ucurv_{\alpha\beta}{}^\mu{}_\nu = 
\partial_\alpha\ucon_{\beta\nu}^\mu +
\ucon_{\alpha\lambda}^\mu\ucon_{\beta\nu}^\lambda - (\alpha\leftrightarrow\beta) = 
\partial_\alpha u_\nu \partial_\beta u^\mu - \partial_\beta u_\nu \partial_\alpha u^\mu
\,.
\end{equation}
One might say that $\ucurv_{\alpha\beta}{}^\mu{}_\nu=0$ means $\ucon^\mu_{\lambda\nu}$ is a ``pure gauge'' connection.

A nontrivial example of flow with $\ucon^\mu_{\lambda\nu}\neq 0$ but
$\ucurv_{\alpha\beta}{}^\mu{}_\nu=0$ is the Bjorken flow. In this case
our proposed choice of $e_a(x)$ in Eq.~(\ref{eq:e-o-e})
 provides a rotationless (i.e., $\mathring\omega=0$) triad field.

\section{A basis in the space orthogonal to $\ph$ and $\bu$ and
  monopole connection}
\label{sec:t_i}

A basis in the space orthogonal to $\ph$ and $\bu$
(cf. Eq. \eqref{eq:tpprojection}) can be obtained easily by rotating
the local confluent basis $e^a$ in such a way that one of the vectors,
say $e_3$, lines up with $\hat q$. The result is given by 
\begin{equation} \label{eq:t_i}
\th^{(i)}
= e^i - (e^3 +\php)\,\frac{\php\cdot e^i}{1+\php\cdot e_3}, \quad
i=1,2,
\end{equation}
satisfying $\th^{(i)}\cdot \th^{(j)} = \delta^{ij}$ and
$\th^{(i)}\cdot\ph_\perp= \th^{(i)}\cdot\bu =0$. 

Since $t^{(i)}$ depends on $x$ (to maintain $\bu(x)\cdot t^{(i)}=0$) as
well as on $\pp$ (to keep $\php\cdot \th^{(i)}=0$), there are two types of
connections in Eq.~(\ref{eq:Nij}) defined by
Eqs.~(\ref{eq:nabla-x-tildeW}) and Eqs.~(\ref{eq:nabla-q}). Applying
these definitions to our choice of $\th^{(i)}$ in
Eq.~(\ref{eq:t_i}) we find for the $x$-derivative connection in Eq.~(\ref{eq:nabla-x-tildeW})
\begin{equation}
  \label{eq:txcon}
 \txcon^{ij}_{\lambda} = \econ^i_{\lambda j} 
- (\econ^i_{\lambda 3}\php\cdot e_j + \econ^3_{\lambda j}\php\cdot e^i)
(1+\php\cdot e_3)^{-1}\,,
\end{equation}
where the connection $\econ^a_{\lambda b}$ is defined by Eq.~(\ref{eq:omega-ede}).

For the $q$-derivative connection, using definition in Eq.~(\ref{eq:nabla-q}),
one obtains
\begin{equation}
	\tcon^{ij}_\mu=\frac{\ph^\lambda(e^i_\lambda
        e^j_\mu-e^j_\lambda e^i_\mu)}{|\pp|+\pp\cdot e_3}
        =\frac{e_3^\lambda(\th^{(i)}_\mu \th^{(j)}_\lambda - \th^{(j)}_\mu \th^{(i)}_\lambda) }{|\pp|+\pp\cdot e_3}
=\varepsilon^{ij}\frac{\varepsilon_{\mu\lambda\nu\sigma}e_3^\lambda
  u^\nu \php^\sigma}{|\pp|+\pp\cdot e_3} \,.
\label{eq:gamma-connection}
\end{equation}
The last expression can be easily recognized as the connection
describing a monopole at $\pp=0$ and Dirac string along
$-e_3$.
The corresponding curvature~\footnote{
Because the space spanned by $t^{(i)}$ is two-dimensional the connection
is abelian, i.e., $[\tcon_\mu,\tcon_\nu]=0$. }
\begin{equation}
  \label{eq:Rmunu}
  \tcurv^{ij}_{\mu\nu} = \partial^{(\p)}_\mu\tcon^{ij}_\nu
  - \partial^{(\p)}_\nu\tcon^{ij}_\mu
  =-\frac{(\th^i_\mu \th^j_\nu - \th^j_\mu \th^i_\nu)}{|\pp|^2} =\varepsilon^{ij}\frac{\varepsilon_{\mu\nu\sigma\lambda} u^\lambda \php^\sigma}{|\pp|^2} \,.
\end{equation}
is the field of a monopole with charge 1 (twice the amount of Berry
curvature monopole charge for spin-$1/2$ fermion). The singularity at
$\pp=0$ is associated with the ambiguity of $\php$ at $\pp=0$.

%%% MS: this is useful for emacs:
%%% Local Variables: 
%%% TeX-PDF-mode: t
%%% TeX-master: "main.tex"
%%% End: 

\section{Comparison to Bjorken flow results}
\label{sec:conservative-fluid}

The purpose of this appendix is to compare our equations with the ones
for a particular case of Bjorken flow derived in Ref. \cite{Akamatsu:2018}.

The first observation we need to make is that the definition of the
equal-time correlator in Ref. \cite{Akamatsu:2018} is subtly different. The Bjorken flow
allows us to define a hypersurface globally which is orthogonal to the
flow 4-vector $ u(x)$ at each point: the constant proper-time
surface $\tau={\rm const}$. It is then natural to define ``equal
time'' correlator in such a way that points $x^\pm$ lie on the same
proper-time hypersurface as $x$. The difference with our definition is
subtle because our equal-time hyperplane is tangential to the
equal-$\tau$ hypersurface at point $x$ and the difference is of order
$y^2$, due to the curvature of the surface.  This difference does lead to
a subtle change in the last term in Eq.~(\ref{eq:kinetic_eq}), which
is necessary to make this equation  agree with Ref. \cite{Akamatsu:2018}.

\begin{figure}
  \centering
  \includegraphics[height=10em]{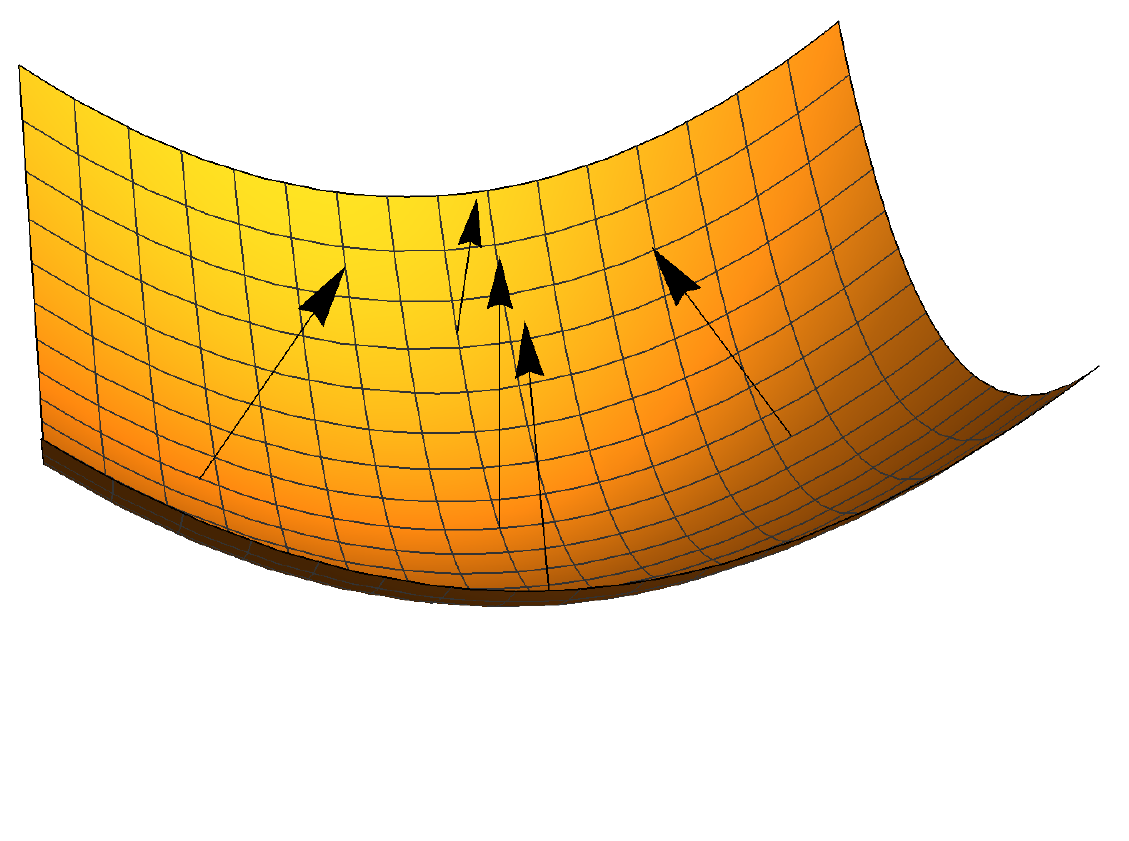}\includegraphics[height=10em]{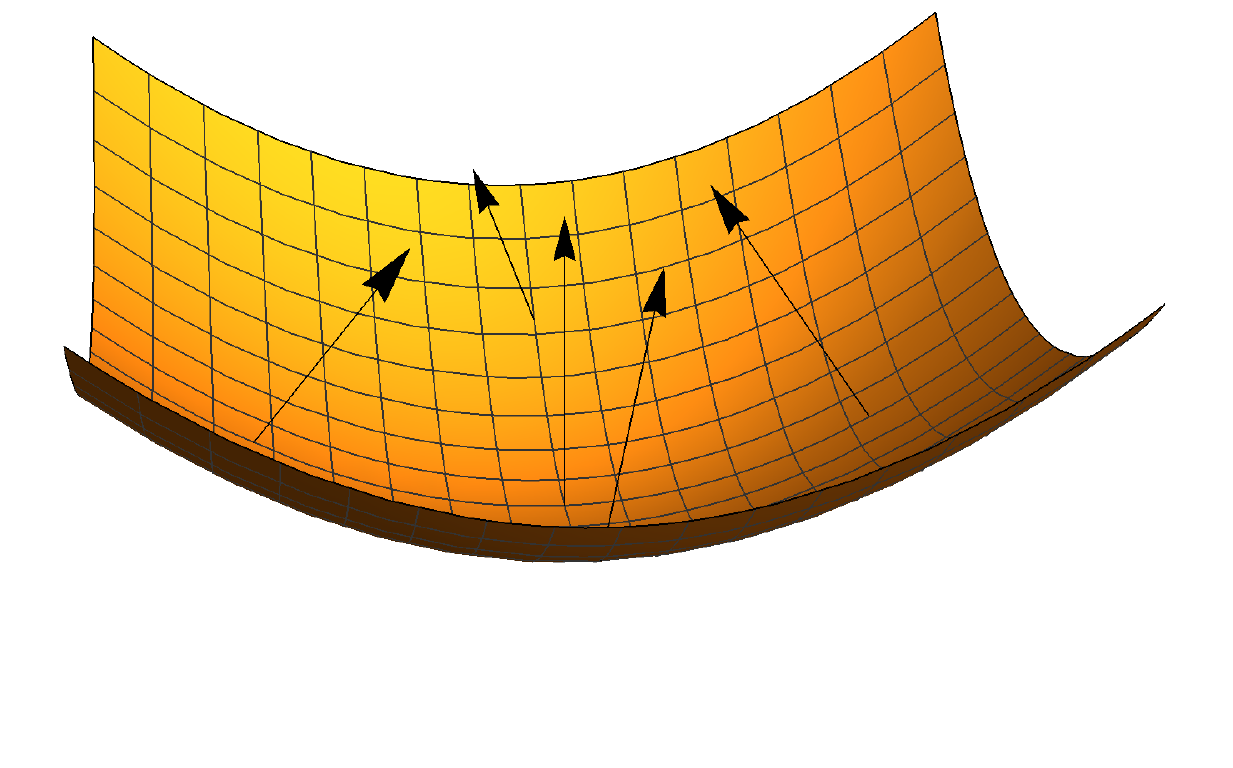}
  \caption{ {\it Left:} Illustration of
    the surface orhogonal to the conservative flow $ u$ at each
    point. Boost is represented by ordinary rotation, preserving
    angles, for clarity. {\it Right:} The same is not possible for
    non-conservative flow, i.e., for nonzero vorticity. However, it is
  possible to make the normal vector to the
  surface (not shown) and the flow vector $ u$ (shown) at the same point be
  different by a purely vortical vector: $v^\mu = \partial^\mu \tau -
  u^\mu$, such that $\partial\cdot v=0$.}
  \label{fig:arrows}
\end{figure}

To describe this in more detail, let us consider a definition of the
equal-time correlator which is slightly different from ours, but will
coinside with $\tau={\rm const}$ for Bjorken flow. It is possible to
define a hypersurface orthogonal to flow if the flow is conservative,
i.e., $ u^\mu = \partial^\mu \tau$ (as is the case for the Bjorken
flow, for example). In general it is not possible, however, one can
perform a Helmholz decomposition into conservative (potential) and
purely vortical flow: $ u^\mu = \partial^\mu \tau + v^\mu$, where
$\partial\cdot v=0$ (see Fig.~\ref{fig:arrows} for illustration). We
will not be interested in doing this {\em globally\/} since we only
need to describe the surface near a given point $x$ to quadratic order
in $y$. Thus we Taylor expand $ u$ to linear order in $\Delta x$:
\begin{equation}
  \label{eq:u-y}
   u_\mu(x+\Delta x) =  u_\mu(x)
  + \frac12(\partial_\mu  u_\nu + \partial_\nu  u_\mu)\Delta x^\nu
  + \frac12(\partial_\mu  u_\nu -  \partial_\nu u_\mu)\Delta x^\nu
\end{equation}
The last term is purely vortical, while the first two terms are
potential, i.e.,
\begin{equation}
  \label{eq:u-Dtau}
  \tau(x+\Delta x) = \tau(x) +  u\cdot \Delta x
  +\frac12 \partial_\mu u_\nu\Delta x^\mu \Delta x^\nu\,.
  \end{equation}
We can then define equal-time correlator in such a way that points $x$
and $x^\pm=x\pm y/2$ lie on the same curved surface $\tau={\rm
  const}$. Using 3-dimensional vector  $y$ in the tangent plane
to $ u(x)$ to parameterize points on such a surface, we can write
explicitly
\begin{equation}
  \label{eq:xpm-14}
  x^\pm_\lambda = x_\lambda \pm \frac{y_\lambda}{2}
  + \frac18 u_\lambda \h_{\mu\nu} y^\mu y^\nu\,,
\end{equation}
where $y\cdot u(x)=0$ and the last term describes the curvature of the surface.
Using this definition of $x^\pm$ instead of Eq. \eqref{eq:xpm}
will change the definition of the "equal-time'' correlator and of the Wigner
function.

In what follows in this section we shall use that modified
definition, but retain the same notation for simplicity.
Due to the modification described above, we must replace $\L^{(y)}$ defined in Eq. \eqref{eq:L_y} by
\begin{eqnarray}
\L^{(y)}&\to&\L^{(y)}-{1\over 4}y^\lambda\Th_\lambda\bu\cdot\del^{(x)},
\end{eqnarray}
where 
\begin{eqnarray}
 \Th_\lambda&\equiv &\frac{c_s}{2}\left( \begin{matrix}
      0 & \h_{\nu\lambda}\\ \h_{\mu\lambda} & 0\\
      \end{matrix}\right).
\end{eqnarray}
As a consequence, Eq. \eqref{eq:kinetic_eq_y} and
\eqref{eq:kinetic_eq} are modified and take the form
\begin{eqnarray}
\bu \cdot \del \N_{AB}(x,y) &=& -\left(\L^{(y)}+{1\over2}{\L}+\Q^{(y)}+\K+\Y\right)_{AC}\, \N^C_{\,\,\,B}(x,y) -\left(-\L^{(y)}+{1\over2}{\L}+\Q^{(y)}+\K+\Y\right)_{BC}\N_A^{\,\,\,C}(x,y) 
\nn
&&+{1\over 4} \big( (y^\lambda\Th_\lambda)_{AC}  \bu\cdot\del^{(x)} \N^C_{\,\,\,B}(x,y)- (y^\lambda\Th_\lambda)_{BC}  \bu\cdot\del^{(x)} \N_A^{\,\,\,C}(x,y) \big)+2 T \bw \Q^{(y)}_{AB} \delta^3(\yp)\nn
\end{eqnarray}
and
\begin{eqnarray}
\bu \cdot \cfd \W(x;\p)&=& -\comm{ i \L^{(\p)}+\K^{(a)}, \W}-  \acomm{{1\over 2} {\Lcon}+\Q^{(\p)}+\K^{(s)}, \W}+ \divu \W+2 T \bw \Q^{(\p)} +(\partial_{\perp\lambda} \bu_\mu) \p_\perp^\mu {\del \W\over \del \p_{\perp \lambda}} 
\nn
&& +{1\over2}a_\lambda \acomm{\left(1-\frac{\dot c_s}{c_s^2}\right)\L^{(\p)},  {\del \W\over \del \p_{\perp \lambda}}}
+{\del \over \del \p_{\perp \lambda}} \left(\{ \Om^{(s)}_\lambda,\W\}+[ \Om^{(a)}_\lambda,\W]-{1\over 4} [\Om_\lambda,[\L^{(\p)},\W]] \right)
\nn
\label{eq:kinetic_eq_conserv}
\end{eqnarray}
respectively, where
 \begin{eqnarray}
   \Om_\lambda\equiv \Hmatrix_\lambda-\Th_\lambda= c_s\left( \begin{matrix}
      0 &  \omega_{\nu\lambda} \\
    \omega_{\mu\lambda} & 0 \\
   \end{matrix}\right).
 \end{eqnarray}
Note, that the only change compared to
Eq.~(\ref{eq:kinetic_eq}) is in the double commutator
term.\footnote{This is consistent with the fact that upon
  diagonalization and time-averaging over faster modes this term
  drops completely. Indeed, the scale of time-averaging, $\ab_t$ is
  much longer than the typical time-like separation between the plane
  tangent to $ u$ and the $\tau={\rm const}$ defined by
  Eq.~(\ref{eq:xpm-14}), which is of order $(\partial u) y^2\sim
  k/q^2\ll 1/q$,
  compared to $\ab_t\gg 1/q$ according to Eq.~(\ref{eq:at}).
}
 For the Bjorken flow, $a_\mu = \omega_{\mu\nu}=0$, and all the terms on
the second line in Eq.~(\ref{eq:kinetic_eq_conserv_Bjorken1})
vanish.

To complete the comparison, for the boost-invariant flow, we perform
the coordinate transformation from $(t, x, y, z)$ to $(\tau, x, y, Y)$
given by $t=\tau\cosh Y,~ x=x,~ y=y,~ z=\tau\sinh Y$, where $\tau$
is the proper time and $Y$ is the space-time rapidity. One can easily
check that  for the Bjorken flow $\bu\cdot\cfd=\partial_\tau$,
$\divu=1/\tau$, $a_\mu=\omega_{\mu\nu}=0$. Thus
Eq. \eqref{eq:kinetic_eq_conserv} is reduced to
\begin{eqnarray}
\partial_\tau \W(x;\p)&=& -\comm{ i \L^{(\p)}+\K^{(a)}, \W}-  \acomm{{1\over 2} {\Lcon}+\Q^{(\p)}+\K^{(s)}, \W}+2 T \bw D^{(\p)} +\frac{1}{\tau}\W+\frac{\p_{ z}}{\tau}  {\del \W\over \del \p_{ z}}.
\label{eq:kinetic_eq_conserv_Bjorken1}
\end{eqnarray}
Since $\p_Y=\tau \p_z$ where $\p_Y$ is the wave vector conjugate to
$Y$, we define $\W_B(x;\p_Y)=\W(x;\p_z)/\tau$ to take into account the
change in the measure of the momentum integration. Using 
\begin{equation}
  \partial_\tau \W(x;\p_z)=\partial_\tau \W(x;\p_z)\Big|_{\p_Y}-\frac{\partial\W(x;\p_z)}{\partial \p_z}(\partial_\tau \p_z)\Big|_{\p_Y}=\partial_\tau\left[\tau\W_B(x;\p_Y)\right]+\frac{ \p_{ z}}{\tau} {\del \W(x;\p_z)\over \del \p_{ z}}
\end{equation}
we obtain
\begin{eqnarray}
\partial_\tau \W_B(x;\p_Y)&=& -\comm{ i \L^{(\p)}+\K^{(a)}, \W_B}-  \acomm{{1\over 2} {\Lcon}+\Q^{(\p)}+\K^{(s)}, \W_B}+\frac{2 T \bw \Q^{(\p)}}{\tau},
\label{eq:kinetic_eq_conserv_Bjorken2}
\end{eqnarray}
where the last two terms in Eq. \eqref{eq:kinetic_eq_conserv_Bjorken1}
were eliminated by the momentum rescaling. Similarly, one can check
that our Eqs.~\eqref{eq:Npm} and \eqref{eq:Nij}, rewritten in terms of
$W_B$, will reduce to Eq. (A7) in Ref. \cite{Akamatsu:2018} exactly.

%%% MS: this is useful for emacs:
%%% Local Variables: 
%%% TeX-PDF-mode: t
%%% TeX-master: "main-v2.tex"
%%% End: 

\section{Notations}
\label{sec:notations}

%This section summarizes our notations.

\begin{list}{}{}
\item
$\cfd_\mu$ -- confluent derivative -- Eqs.~(\ref{eq:nabla-phi-A}),~(\ref{eq:nabla-mu-wigner}),~(\ref{eq:nabla-x-tildeW});
\item
$\gamma$ -- generic relaxation constant used in order-of-magnitude expressions;
\item
$\gamma_\eta$, $\gamma_\zeta$, $\gamma_L$ -- shear, bulk and
longitudinal (sound) relaxation constants Eqs.~(\ref{eq:gammas})
and~(\ref{eq:gammaL});
\item
$\Delta^{\mu\nu}$ -- projector on hyperplane orthogonal to $u$ --
Eq.~(\ref{eq:Delta-definition});
%\item
%$\Delta^{\mu\nu}$ -- projector on hyperplane orthogonal to $\bu$ --
%Eq.~(\ref{eq:Delta-definition});
\item
$\delta e$, $\delta g^\mu$ -- scaled fluctuations of energy density (times $c_s$) and
velocity (times $ w$)  -- Eq.~(\ref{eq:phi_defn});
\item
$\eps$, $u$ -- local averaged energy density and fluid velocity -- Eq.~(\ref{eq:du-deps});
\item
$\seps$, $\su$ -- fluctuating energy density and fluid velocity -- Eq.~(\ref{eq:du-deps});
\item
$\theta^{\mu\nu}$ -- symmetrized velocity gradients -- Eq.~(\ref{eq:theta-def});
\item 
$\Lambda$ -- with no argument -- wave-vector cutoff, $\Lambda = 1/b$;
\item 
$\Lambda(\Delta x)$ -- Lorentz boost bringing fluid at point
$x+\Delta x$ to rest with respect to fluid at point $x$ -- Eq.~(\ref{eq:u-Lambda-u});
\item
$\phi_A$ -- ($A=e,0,1,2,3$) the set of fluctuations of hydrodynamic
variables -- Eq.~(\ref{eq:phi_defn});
\item
$\psi_A$ -- the set of hydrodynamic variables, e.g., $(\epsilon,u^\mu)$;
\item
$\omega^{\mu\nu}$ -- vorticity -- Eq.~(\ref{eq:omega-def});
\item 
$\ucon^\nu_{\lambda\mu}$ -- confluent connection -- Eq.~(\ref{eq:Gamma-udu});
\item 
$\econ^a_{\mu b}$ -- spin connection for local triad $e^a$ -- Eq.~(\ref{eq:omega-ede});
\item
$\tcon^{ij}_\mu$ -- momentum space spin connection for diad $t^{(i)}$
-- Eq.~(\ref{eq:nabla-q});
\item
$\txcon^{ij}_\mu$ -- coordinate space spin connection for diad $t^{(i)}$
-- Eq.~(\ref{eq:nabla-x-tildeW});
\item
$a^\mu$ -- local acceleration $a^\mu =( u\cdot\partial) u^\mu$;
\item 
$\ab$ -- coarse-graining scale -- Eq.~(\ref{eq:a-vs-lmic});
\item 
$\ab_t$ -- temporal coarse-graining scale -- Eq.~(\ref{eq:at});
\item
$\dot c_s$ -- logarithmic rate of dependence of sound speed on entropy
-- Eq.~(\ref{eq:csdot});
\item
$\partial_\perp$ -- partial derivative projected on hyperplane
orthogonal to 4-velocity -- Eq.~(\ref{eq:partial_perp});
\item
$\partial_\mu \W_{AB}(x,q)$ -- partial $x$-derivative at fixed 
$q^a=e^a(x)\cdot q$ -- Eq.~(\ref{eq:nabla-mu-wigner}); 
\item
  $E$ or $E(\pq_\perp)$ -- phonon energy -- Eq.~(\ref{eq:f(p)});
\item $e^a_\mu$ -- ($a=1,2,3$) local triad basis vector orthogonal
  to $ u(x)$ --
  Eq.~(\ref{eq:nabla-a}), Appendix~\ref{sec:tetrad};
\item
  $F_\pm$ -- Eq.~(\ref{eq:FF});
\item
$\N_{AB}(x)\equiv \N_{AB}(x,0)$ -- correlator at coincident points --
Eq.~(\ref{eq:G_AB-phi-phi});
\item 
$\wt\N_{AB}(x)$ -- finite part of  $\N_{AB}(x)$ 
-- Eq.~(\ref{eq:Gtilde});
\item
$\N_{AB}(x,y)$ -- equal-time correlator -- Eq.~(\ref{eq:GAB});
\item
$\GG_{AB}(x,y)$ -- confluent equal-time correlator -- Eq.~(\ref{eq:LGL});
\item
$k$ -- wave vector Fourier conjugate to midpoint vector $x$;
\item
$\WN(x,p)$ -- phonon phase-space distribution function -- Eq.~(\ref{eq:Nhat-N});
\item
$\WN_\pm(x,p)$ -- phase-space distribution function for
positive/negative frequency phonons;
\item
$\pq$ -- phonon momentum (not to be confused with pressure $p(\eps)$) -- Section~\ref{sec:phonon};
\item
$ p\equiv p(\eps)$ -- pressure at average energy density $\eps$;
\item
$\p$ -- wave vector Fourier conjugate to separation vector $y$, also $q\cdot\bu=0$ -- Section~\ref{sec:covar-deriv-conn};
\item 
$T$ -- temperature (local value at $\eps(x)$);
\item
$t^{(i)}_\mu$ -- ($i=1,2$) momentum space diad vector orthogonal to $\p^\mu$
-- Eq.~(\ref{eq:ti}), Appendix~\ref{sec:t_i};
\item
$ u^\mu(x)$ -- local averaged 4-velocity at point $x$; 
\item
$ w\equiv\eps+p(\eps)$ -- enthalpy density at average energy density;
\item
$W_{AB}(x,\p)$ -- Wigner function, Wigner/Fourier transform of
$\GG_{AB}(x,y)$ -- Eq.~(\ref{eq:Wigner-function});
\item
$\wt W_{AB}(x,\p)$ -- $W_{AB}(x,\p)$ after subtraction of leading and
subleading large-$q$ terms -- Eq.~(\ref{eq:tildeW-W0W1}).
\item
$W_{\cA\cB}(x,\p)$  -- Wigner function $W_{AB}(x,\p)$ in the basis of
ideal hydrodynamics modes -- Eqs.~(\ref{eq:eigenvectors}),~(\ref{eq:W_bAbB});
\item
$W_\pm\equiv W_{\pm\pm}$ --  Wigner functions for positive/negative frequency sound modes
-- Eqs.~(\ref{eq:eigenvectors}),~(\ref{eq:W_bAbB});
\item
$\wtr^{ij}$ -- $2\times2$ matrix of Wigner functions for transverse modes -- Eq.~(\ref{eq:Nij});
\item
$x$ -- in a two-point correlator -- the midpoint space-time vector;
\item
$x^\pm$ -- arguments of the 2-point correlator;
\item 
$y$ -- in an equal-time two-point correlator -- the separation
  vector, constrained by $ u(x)\cdot y=0$.
\end{list}

%%% MS: this is useful for emacs:
%%% Local Variables: 
%%% TeX-PDF-mode: t
%%% TeX-master: "main.tex"
%%% End:

\bibliographystyle{utphys}
\bibliography{references}

\end{document}